\pdfoutput=1

\documentclass[fleqn,usenatbib]{mnras}

\usepackage{txfonts} 
\usepackage[T1]{fontenc}
\usepackage{threeparttable}
\usepackage{amssymb}
\usepackage{graphicx}
\usepackage{float}
\usepackage{morefloats}
\usepackage[usenames,dvipsnames,svgnames,table]{xcolor}
\usepackage{multirow}
\usepackage{caption}
\usepackage[all]{hypcap}
\hypersetup{
    colorlinks,
    linkcolor={red!50!black},
    citecolor={blue!50!black},
    urlcolor={blue!80!black}
}

\graphicspath{{figs/}}

\title[Mapping CRRLs towards Cassiopeia~A]{Mapping low frequency carbon radio recombination lines towards Cassiopeia~A at $340$, $148$, $54$ and $43$~MHz}

\author[P.~Salas et al.]{
P.~Salas$^{1}$\thanks{E-mail: psalas@strw.leidenuniv.nl}, 
J.~B.~R. Oonk$^{1,2}$, 
R.~J.~van~Weeren$^{3}$,
M.~G.~Wolfire$^{4}$,
K.~L.~Emig$^{1}$,\newauthor
M.~C.~Toribio$^{1}$,
H.~J.~A.~R\"ottgering$^{1}$ and 
A.~G.~G.~M.~Tielens$^{1}$\\
$^{1}$Leiden Observatory, Leiden University, P.O. Box 9513, NL-2300 RA Leiden, The Netherlands\\
$^{2}$Netherlands Institute for Radio Astronomy (ASTRON), Postbus 2, 7990 AA Dwingeloo, The Netherlands\\
$^{3}$Harvard-Smithsonian Center for Astrophysics, 60 Garden Street, Cambridge, MA 02138, USA\\
$^{4}$Department of Astronomy, University of Maryland, College Park, MD 20742-2421, USA\\
}

\date{Accepted XXX. Received YYY; in original form ZZZ}

\begin{document}

\label{firstpage}
\pagerange{\pageref{firstpage}--\pageref{lastpage}}
\maketitle

\begin{abstract}
Quantitative understanding of the interstellar medium requires knowledge of its physical conditions.
Low frequency carbon radio recombination lines (CRRLs) trace cold interstellar gas, and can be used to determine its physical conditions (e.g., electron temperature and density).
In this work we present spatially resolved observations of the low frequency ($\leq390$~MHz) CRRLs centered around C$268\alpha$, C$357\alpha$, C$494\alpha$ and C$539\alpha$ towards Cassiopeia~A on scales of $\leq1.2$~pc.
We compare the spatial distribution of CRRLs with other ISM tracers.
This comparison reveals a spatial offset between the peak of the CRRLs and other tracers, which is very characteristic for photodissociation regions and that we take as evidence for CRRLs being preferentially detected from the surfaces of molecular clouds.
Using the CRRLs we constrain the gas electron temperature and density.
These constraints on the gas conditions suggest variations of less than a factor of two in pressure over $\sim1$~pc scales, and an average hydrogen density of $200$--$470$~cm$^{-3}$.
From the electron temperature and density maps we also constrain the ionized carbon emission measure, column density and path length.
Based on these, the hydrogen column density is larger than $10^{22}$~cm$^{-2}$, with a peak of $\sim4\times10^{22}$~cm$^{-2}$ towards the South of Cassiopeia~A.
Towards the southern peak the line of sight length is $\sim40$~pc over a $\sim2$~pc wide structure, which implies that the gas is a thin surface layer on a large (molecular) cloud that is only partially intersected by Cassiopeia A.
These observations highlight the utility of CRRLs as tracers of low density extended HI and CO-dark gas halo's around molecular clouds.
\end{abstract}

\begin{keywords}
ISM: clouds -- radio lines : ISM -- ISM: individual objects: Cassiopeia A
\end{keywords}

\section{Introduction}
\label{sec:intro}

Molecular hydrogen, the material that fuels star formation, is formed out of atomic hydrogen \citep[e.g.,][]{Cazaux2004}.
This is clear in the interstellar medium (ISM), where we observe that molecular gas is embedded in atomic hydrogen \citep[e.g.,][]{Andersson1991,Williams1996,Moriarty-Schieven1997,Fukui2009,Pascucci2015}.
Despite the clear association between these two gas compositions, the exact details of what is this atomic envelope are not clear \citep[e.g.,][]{Blitz1999,Hollenbach1999}.
In order to understand better the relation between mostly molecular dense gas and mostly atomic diffuse gas a larger sample of gas in this transition regime is required, along accurate estimates of its temperature and density.

A way in which we can study the cold atomic gas in the envelopes of molecular clouds is through observations of low frequency ($\nu\lesssim1$~GHz) carbon radio recombination lines \citep[CRRLs, e.g.,][]{Gordon2009}.
The population of carbon ions recombining to a given principal quantum number, $n$, is determined by the gas density, temperature and radiation field, as well as the atomic physics involved \citep[e.g.,][]{Shaver1975,Watson1980,Salgado2017a}. 
Thus, we can determine the gas physical conditions by comparing the observed properties of CRRLs at different frequencies with model predictions \citep[e.g.,][]{Payne1989,Kantharia1998,Oonk2017}.
Using this method it has been determined that CRRLs trace cold ($T\sim100$~K) diffuse gas \citep[$n_{\rm{H}}\sim100$~cm$^{-3}$, e.g.,][]{Konovalenko1984,Ershov1987,Sorochenko1991,Payne1994,Kantharia1998,Roshi2011,Oonk2017} in the Galaxy.
These physical conditions are similar to those found from observations of atomic hydrogen associated with molecular clouds, either using self absorption features 
\citep[HISA e.g.,][]{Gibson2002,Kavars2003,Kerton2005,Kavars2005,Moss2012} or absorption measurements against bright background continuum sources \citep[HICA e.g,][]{Dickey2009,Stanimirovic2014,Bihr2015}.

Most of our understanding of low frequency CRRLs in the Galaxy comes from studies where the spatial resolution is coarse \citep[e.g., $\theta_{\rm{HPBW}}\gtrsim30\arcmin$][]{Anantharamaiah1988,Erickson1995,Kantharia2001,Roshi2002}.
This has hindered a spatially resolved identification of which components of the ISM do low frequency CRRLs preferentially trace, such as the outskirts of HII regions, diffuse CII clouds and/or the envelopes of molecular clouds.
A notable exception to this limitation is the line of sight towards the supernova remnant Cassiopeia~A (Cas~A).
Along this line of sight, the bright background continuum (relative to the diffuse synchrotron emission from the Milky Way) enables RRL studies with an effective resolution comparable to the size of Cas~A \citep[diameter of $6\arcmin$ at $64$~MHz, e.g.,][]{Oonk2017}.
Additionally, the large gas column density in the intervening ISM has allowed a direct detection of CRRLs with a spatial resolution smaller than the size of Cas~A \citep{Anantharamaiah1994,Kantharia1998,Asgekar2013}.
From these observations we know that the optical depth of the CRRLs associated with gas in the Perseus arm increases towards the South of Cas~A \citep{Anantharamaiah1994,Kantharia1998}, peaking against its western hotspot \citep{Asgekar2013}.

\begin{figure*}
 \begin{center}
  \includegraphics[width=0.95\textwidth]{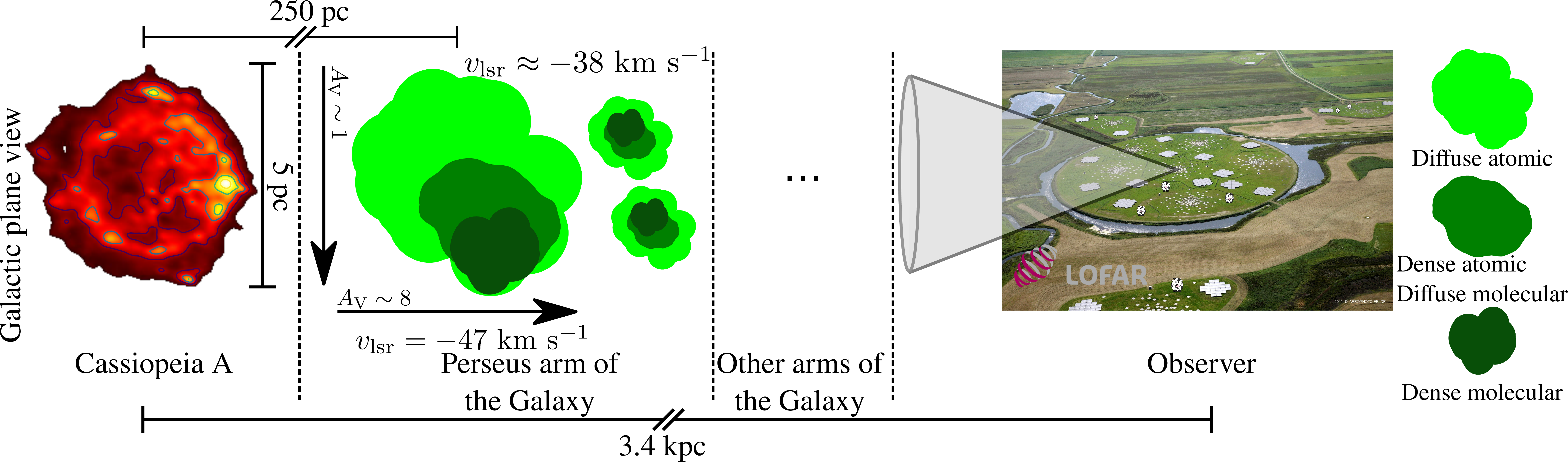}
 \end{center}
 \caption{Cartoon depicting the geometry of the studied gas relative to Cassiopeia~A and the observer. 
 The gas studied in this work is part of the Perseus arm of the Galaxy and does not show signs of being related to the supernova remnant.
 The dense molecular gas peaks to the south of Cas~A, as shown by continuum observations in the far infrared \citep{deLooze2017} and CO lines \citep[e.g.,][]{Liszt1999,Kilpatrick2014}. 
 Here we only illustrate the gas over the face of Cas~A as our RRL observations do not trace gas outside the face of Cas~A.}
 \label{fig:cascartoon}
\end{figure*}

Besides low frequency CRRLs, the line of sight towards Cas~A has been the target of numerous studies of the ISM \citep[e.g.,][]{Davies1972,Mebold1975,Troland1985,Bieging1986,Wilson1993,Liszt1999,deLooze2017}.
Given that Cas~A is located on the far side of the Perseus arm of the Galaxy \citep[at a distance of $3.4$~kpc from the observer][]{Reed1995}, most of the Perseus arm gas lies between the observed and the background source \citep[e.g.,][]{Troland1985}.
Additionally, the distance between Cas~A and the gas in the Perseus arms is large enough that they should be unrelated \citep[][]{Xu2006,Sorochenko2010,Choi2014,Salas2017}.
The spatial distribution of the atomic gas towards Cas~A shows that it completely covers the face of Cas~A, as revealed by observations of the $21$~cm line of HI in absorption against Cas~A \citep[e.g.,][]{Bieging1991,Schwarz1997}.
However, the saturation of the $21$~cm-HI line profiles makes it difficult to identify small scale structure.
Observations of other tracers have revealed the presence of gas with a large column density over the southern half of Cas~A, with a visual extinction in the range $A_{\rm{V}}\sim4\mbox{--}10$~\citep[e.g.,][]{Troland1985,Hwang2012,deLooze2017}.
A cartoon illustrating the distribution of the gas relative to Cas~A and the observer is shown in Figure~\ref{fig:cascartoon}.

The larger optical depth of CRRLs towards the south of Cas~A \citep{Anantharamaiah1994,Kantharia1998}, where CO emission is readily detected, provides evidence, for the association of low frequency CRRLs with cold atomic or diffuse molecular envelopes of molecular clouds \citep[e.g.,][]{Andersson1991}.
Another argument in favor of this association comes from the gas temperature and density as traced by low frequency CRRLs.
Using spatially unresolved observations of CRRLs \citet{Oonk2017} derived a gas electron temperature of $85$~K and a density of $\sim280$~cm$^{-3}$, in between those of cold molecular and diffuse atomic gas.
A step forward in this direction would be to extend this spatial comparison to the gas physical conditions, which was not possible previously due to the lack of frequency coverage. 

In this work we present CRRL emission and absorption cubes centred around the C$268\alpha$, C$357\alpha$ C$494\alpha$ and C$539\alpha$ lines with a resolution of $70\arcsec$ ($1.2$~pc at the distance of Cas~A).
With these cubes we aim to study the relation between the gas traced by low frequency CRRLs and other tracers of cold gas such as cold atomic gas traced by the $21$~cm line of HI in absorption and molecular gas as traced by CO lines in the millimetre.
We perform this comparison both spatially and in terms of the physical conditions as derived from two sets of lines; one containing the CRRLs and the second the molecular lines.

\section{Observations \& data reduction}
\label{sec:obs}

Here we describe the data reduction of the low frequency array \citep[LOFAR,][]{vanHaarlem2013} high band antenna (HBA) observations presented in this work, as well as further processing steps applied to the previously published observations used.
For the details about the observations collected from the literature we refer the reader to the original works (Table \ref{tab:obs}).

\begin{table*}
\begin{center}
\begin{threeparttable}
\caption{Selected observations towards Cas~A}
\begin{tabular}{llccc}
\hline
\hline
Line or band & Telescope & Velocity resolution & Spatial resolution\tnote{c} & Reference\\
             &           & (km~s$^{-1}$)       &                             &          \\
\hline
C$268\alpha$                     & WSRT & $0.6$  & $70\arcsec\times70\arcsec$ $0\degr$   & This work. \\
C$357\alpha$                & LOFAR HBA & $0.7$  & $18\arcsec\times18\arcsec$ $0\degr$   & This work. \\
C$494\alpha$                & LOFAR LBA & $2$    & $65\arcsec\times45\arcsec$ $100\degr$ & \cite{Oonk2017} \\
C$539\alpha$                & LOFAR LBA & $2$    & $65\arcsec\times45\arcsec$ $100\degr$ & \cite{Oonk2017} \\
HI--$21$~cm                         & VLA  & $0.65$ & $7\arcsec\times7\arcsec$ $0\degr$     & \cite{Bieging1991} \\
$1667$~MHz--OH                      & VLA  & $1.3$  & $7\arcsec\times7\arcsec$ $0\degr$     & \cite{Bieging1986} \\
$^{13}$CO$(1\mbox{--}0)$\tnote{a}  & NRAO $12$~m & $0.13$ & $56\arcsec\times56\arcsec$ $0\degr$   & \cite{Liszt1999} \\
CO$(2\mbox{--}1)$\tnote{b}         & SMT  & $0.3$  & $33\arcsec\times33\arcsec$ $0\degr$   & \cite{Kilpatrick2014} \\
~[CI] $^{3}$P$_{1}$--$^{3}$P$_{0}$ & KOSMA & $0.63$ & $54\arcsec\times54\arcsec$ $0\degr$  & \cite{Mookerjea2006} \\
~[CII]          & Herschel PACS\tnote{d} & $250$ & $12\arcsec\times12\arcsec$ $0\degr$              & \cite{Salas2017} \\
~$A_{\rm{V}}$          &   &    & $35\arcsec\times35\arcsec$ $0\degr$              & \cite{deLooze2017} \\
\hline
\end{tabular}
\begin{tablenotes}
      \small
      \item[a] $^{12}$CO$(2\mbox{--}1)$ is also available from the same observations.
      \item[b] $^{12}$CO and $^{13}$CO.
      \item[c] Observing beam major axis, minor axis and position angle.
      \item[d] {\footnotesize Herschel is an ESA space observatory with science instruments provided by European-led Principal Investigator consortia and with important participation from NASA.}
\end{tablenotes}
\label{tab:obs}
\end{threeparttable}
\end{center}
\end{table*}

\subsection{WSRT data}

The WSRT data used in this work is the same presented in \citet{Oonk2017}.
The data reduction steps are the same up to the imaging part.
Before imaging, we subtracted the continuum from the calibrated visibilities using \emph{CASA}'s \citep{McMullin2007} \emph{uvcontsub} \citep[see e.g.,][]{vanGorkom1989,vanLangevelde1990}.
We use a first order polynomial which is fit to line free channels at both sides of the $\alpha$ lines when possible.
After this we imaged the continuum subtracted data using Briggs weighting \citep{Briggs1995}.
We tested using different robust parameters to determine the best trade-off between sensitivity and resolution.
Using a robust factor of $-1$ provided the best angular resolution $\theta\leq70\arcsec$ for most subbands ($40$ out of the $6\times8$ subbands).
A robust factor of $1$ provides a factor of two lower spectral noise with a synthesised beam size of $\theta\leq100\arcsec$ ($47/48$ subbands).
We use the cubes generated with a robust of $-1$ since we are interested in the spatial structure of the line.
After imaging we stack the cubes and apply a bandpass correction in the image plane.
The stacked cube has a spatial resolution of $70\arcsec$ and contains $40$ $\alpha$ RRL transitions\footnote{A line involving a change in principal quantum number of $\Delta n=1$ is called an $\alpha$ line.}.
The stacked line profile corresponds to RRLs with an average $n$ of $268$.
After this we compared the stacked line profile extracted from over the face of Cas~A with those presented by \citet{Oonk2017}.
This comparison showed that the spectra agree within errors.

To study the distribution of the weaker C$268\alpha$ $-38$~km~s$^{-1}$ velocity components we convolve the spectral axis of the cube to increase the signal-to-noise.
As convolution kernel we use a boxcar four channels wide.
This also produces a cube with a similar velocity resolution as that of the C$539\alpha$ cube (see Table~\ref{tab:obs}).
In order to allow for a better comparison between the C$268\alpha$ and C$539\alpha$ lines we regrid the spectral axis of the C$268\alpha$ cube to match that of the C$539\alpha$ cube.

Of the CRRL data used in this work, the one coming from the WSRT observations is the one with the coarsest spatial resolution.

\subsection{LOFAR HBA data}

Cas~A was observed with the LOFAR HBA on December $13$, $2015$ for four hours (obsid: L415239).
This data was taken as part of the LOFAR Cassiopeia~A spectral survey (LCASS, PI, J.~B.~R.~Oonk).
During the observation all $23+14$ Dutch stations were used.
These observations cover the $132.6$--$152.3$~MHz range with $195.3125$~kHz wide spectral windows.
The correlator was set up to deliver spectral windows with $512$ spectral channels.
This results in a channel width of $0.75$--$0.87$~km~s$^{-1}$.

Cygnus~A was used as amplitude calibrator at the beginning of the observations (obsid: L415237).
Phase and amplitude solutions were derived against Cygnus~A and then applied to the Cas~A data.
After transferring the amplitude and phase from Cygnus~A, we self calibrated the Cas~A data.
We started the self-calibration cycle using a small number of clean iterations and short baselines, then in each repetition of the cycle a larger number of clean iterations, as well as longer baselines, were used.
The cut-off in the baseline length started at $2000$~lambdas and increased to $12000$~lambdas.
LOFAR has baselines longer than $12000$~lambdas, but we decided to stop at this cut-off because the signal-to-noise ratio drops for higher resolution.
After imaging, the cubes were convolved to a common resolution of $18\arcsec$.
The cubes were then converted to optical depth using $\tau_{\nu}=I_{\nu}/I_{\nu}^{\rm{cont}}-1$, where $I_{\nu}$ is the spectrum extracted from the data cubes and $I_{\nu}^{\rm{cont}}$ is the continuum determined from a linear fit to line free channels \citep[e.g.,][]{Oonk2014,Salas2017}.
Any residual bandpass in the optical depth cubes was corrected in the image plane using an order two polynomial.

After this, the RRL optical depth cubes were stacked.
In the frequency range between $132.6$ and $152.3$~MHz there are $16$ $\alpha$ RRLs.
From these, we selected four lines which were in spectral windows with low radio frequency interference during the observations.
The stacked cube has $\alpha$ RRLs with an averaged principal quantum number $n=357$.

\subsection{LOFAR LBA data}

The data reduction of the LOFAR LBA data is described in \citet[][obsid $40787$]{Oonk2017}.
For this work we have split the $\sim75$ CRRLs present in the $33$--$57$~MHz range into two groups;
\setcitestyle{notesep={ }}
one group uses the first six CRRLs, which have principal quantum numbers $n=491,492,493,495,496,497$, and another group with the remaining CRRLs \citep[see Table~2 of][for the complete list]{Oonk2017}.
\setcitestyle{notesep={, }}
This division is made in order to study the CRRL profile at a lower $n$ number, where the effects of pressure and radiation broadening are less severe and it is easier to differentiate velocity components \citep[e.g.,][]{Oonk2017}.
The stacked cubes have $\alpha$ RRLs with averaged principal quantum numbers of $494$ and $539$.

\subsection{Literature data}

We complement the spatially resolved LOFAR and WSRT cubes presented in this work with observations from the literature. 
A summary of the literature observations is presented in Table~\ref{tab:obs}.
From the literature we have selected the following maps; HI--$21$~cm \citep{Bieging1991}, the $1667$~MHz line of OH \citep{Bieging1986}, $^{12}$CO(2--1) and $^{13}$CO(2--1) \citep{Kilpatrick2014}, $492$~GHz [CI] \citet{Mookerjea2006}, and $158$~$\mu$m [CII] \citep{Salas2017}.
Additionally we include the dust-derived interstellar extinction A$_{\rm{V}}$ map of \citet{deLooze2017}.

To compare observations with different angular resolutions we convolve the maps to a common resolution of $70\arcsec$.
We use $70\arcsec$ to match the resolution of the WSRT cubes.
Two exceptions are the $158$~$\mu$m [CII] cube and A$_{\rm{V}}$ map.
We do not convolve the $158$~$\mu$m [CII] cube since the area covered by each PACS observation is smaller ($45\arcsec\times45\arcsec$) than the target resolution.
As for the dust-derived A$_{\rm{V}}$ map, we do not convolve because the images used to model the dust emission were analysed at $35\arcsec$ resolution \citep{deLooze2017}.

\section{Results}
\label{sec:results}

\subsection{Global velocity structure}

In terms of the line of sight structure, the gas towards Cas~A is observed in various ISM tracers in at least four velocity components.
One component corresponds to gas in the local Orion spur at velocities close to $0$~km~s$^{-1}$ (all velocities are referenced with respect to the local standard of rest).
The remaining velocity components, and main focus of this work, are associated with the Perseus arm of the Galaxy at velocities of $-47$, $-41$ and $-36$~km~s$^{-1}$.
These last two velocity components have been treated as a single velocity component at $-38$~km~s$^{-1}$ in previous CRRL studies because they are difficult to separate \citep[e.g.,][]{Payne1994,Kantharia1998,Oonk2017}.
Here we use this nomenclature when we are not able to separate the $-41$ and $-36$~km~s$^{-1}$ velocity components.

\begin{figure}
 \begin{center}
  \includegraphics[width=0.49\textwidth]{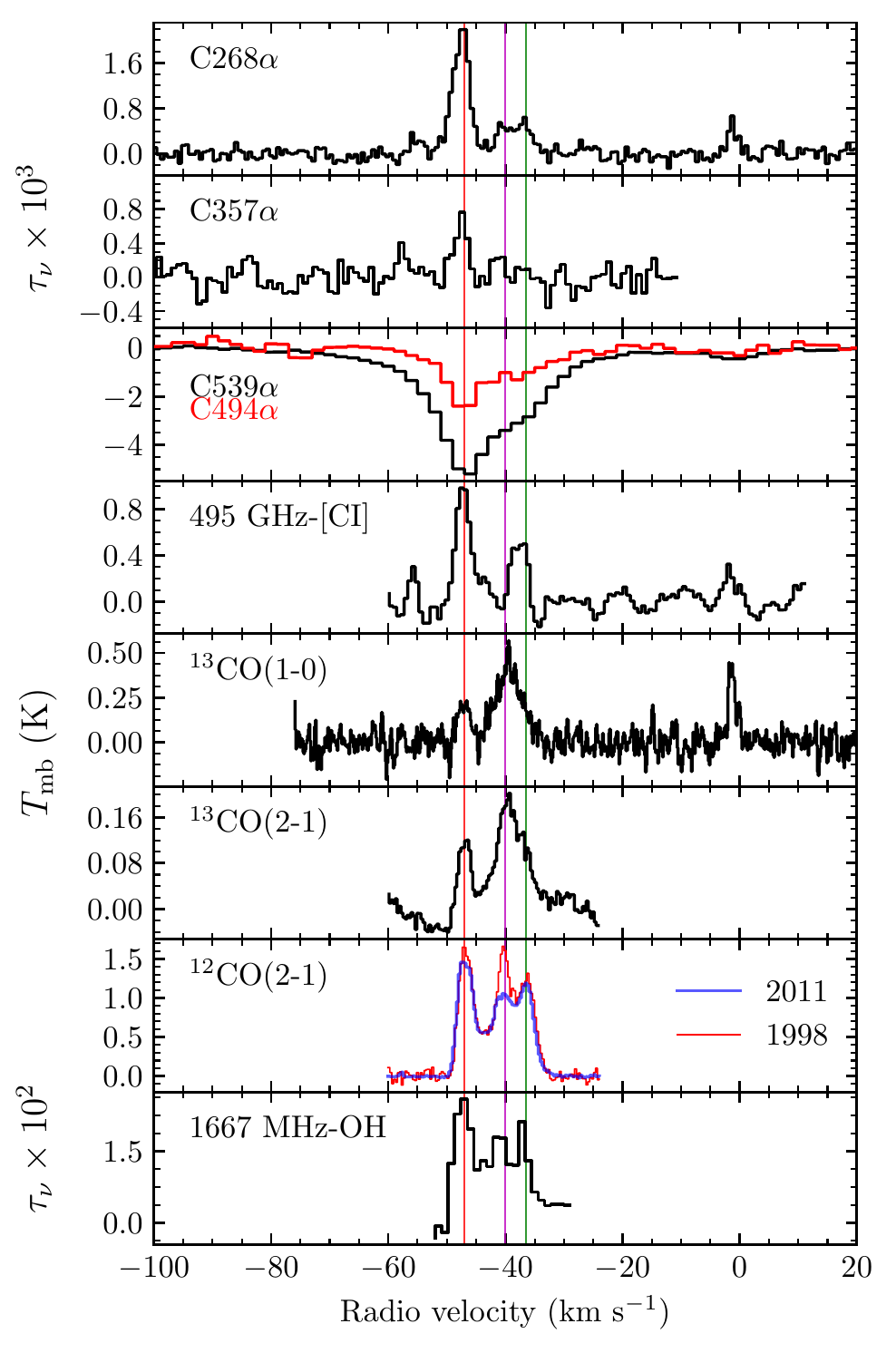}
 \end{center}
 \caption{Comparison between the spectra of C$268\alpha$, C$357\alpha$, C$539\alpha$, $492$~GHz--[CI], $^{13}$CO$(1$--$0)$, $^{13}$CO$(2$--$1)$, $^{12}$CO$(2$--$1)$ and $1667$~MHz--OH.
 The three CRRLs and the OH line are shown in optical depth units while the [CI] and CO lines in brightness temperature units.
 The spectra where extracted from an aperture defined by the extent of the $1667$~MHz--OH line map.
 The Perseus arm velocity components at $-47$, $-41$ and $-36.5$~km~s$^{-1}$ are shown with {\it red}, {\it magenta} and {\it green} lines respectively.
 The difference in the $1998$ and $2011$ $^{12}$CO$(2$--$1)$ line profiles is due to the different choice of off-source position.}
 \label{fig:specavg}
\end{figure}

To compare the line profiles we averaged the pixels covering the face of Cas~A.
We define the face of Cas~A as a circle of radius $2\farcm5$ centred on $(\alpha,\delta)_{\mbox{J2000}}=(23^{\mbox{h}}23^{\mbox{m}}24^{\mbox{s}},+58^{\circ}48\arcmin54\arcsec)$.
The spectra are shown in Figure~\ref{fig:specavg}.
In this Figure we highlight the position of three velocity components, at $-47$, $-41$ and $-36.5$~km~s$^{-1}$.
When the spectra shows the presence of these three velocity components, we notice that the velocity of the line peak agrees between different tracers.
For the $-38$~km~s$^{-1}$ velocity component we can see that the line profile is a blend of two or more velocity components.
This is more readily seen in the line profiles of C$268\alpha$, $^{12}$CO and $1667$~MHz--OH, where two velocity components are observed, one close to $-41$~km~s$^{-1}$ and other at $-36.5$~km~s$^{-1}$.

\subsection{Channel maps}

\begin{figure}
 \begin{center}
  \includegraphics[width=0.49\textwidth]{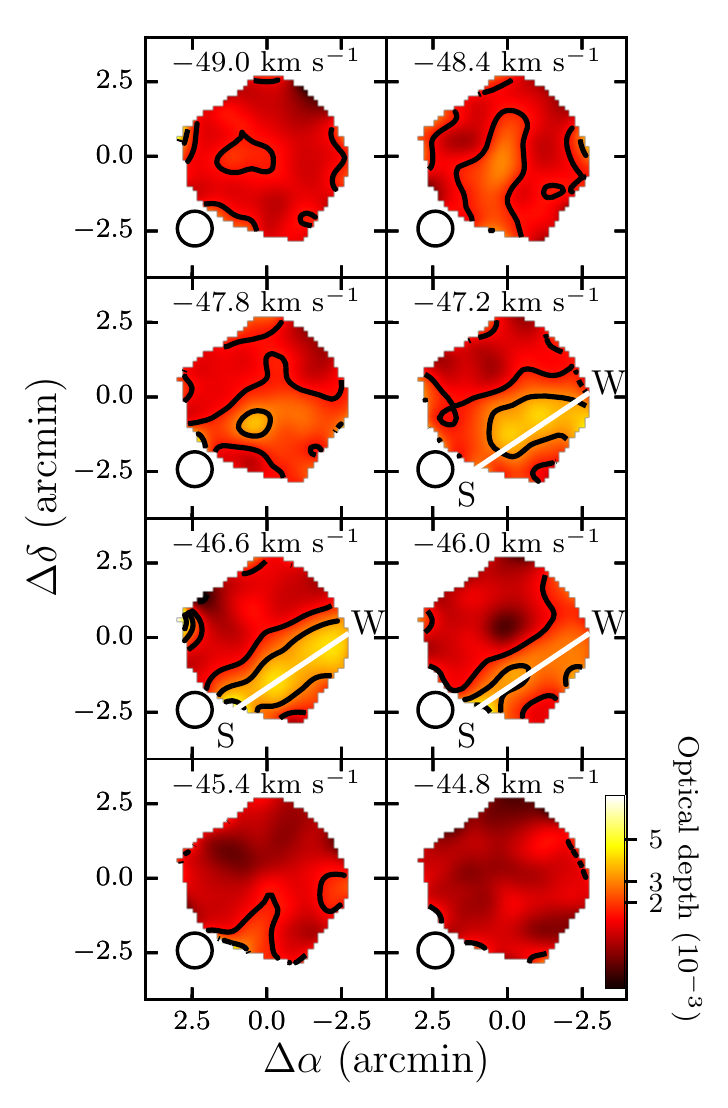}
 \end{center}
 \caption{C$268\alpha$ emission around $-47$~km~s$^{-1}$ channel maps.
 These maps have a spatial resolution of $70\arcsec$, shown in the lower left corner of each panel.
 The contours start at $3\sigma$, with $\sigma=6\times10^{-4}$, and continue at values of $4\sigma$, $5\sigma$ and $6\sigma$.
 The white line marks the location of an elongated structure with a W-S orientation.}
 \label{fig:chwsrt}
\end{figure}

\begin{figure*}
 \begin{center}
  \includegraphics[width=\textwidth]{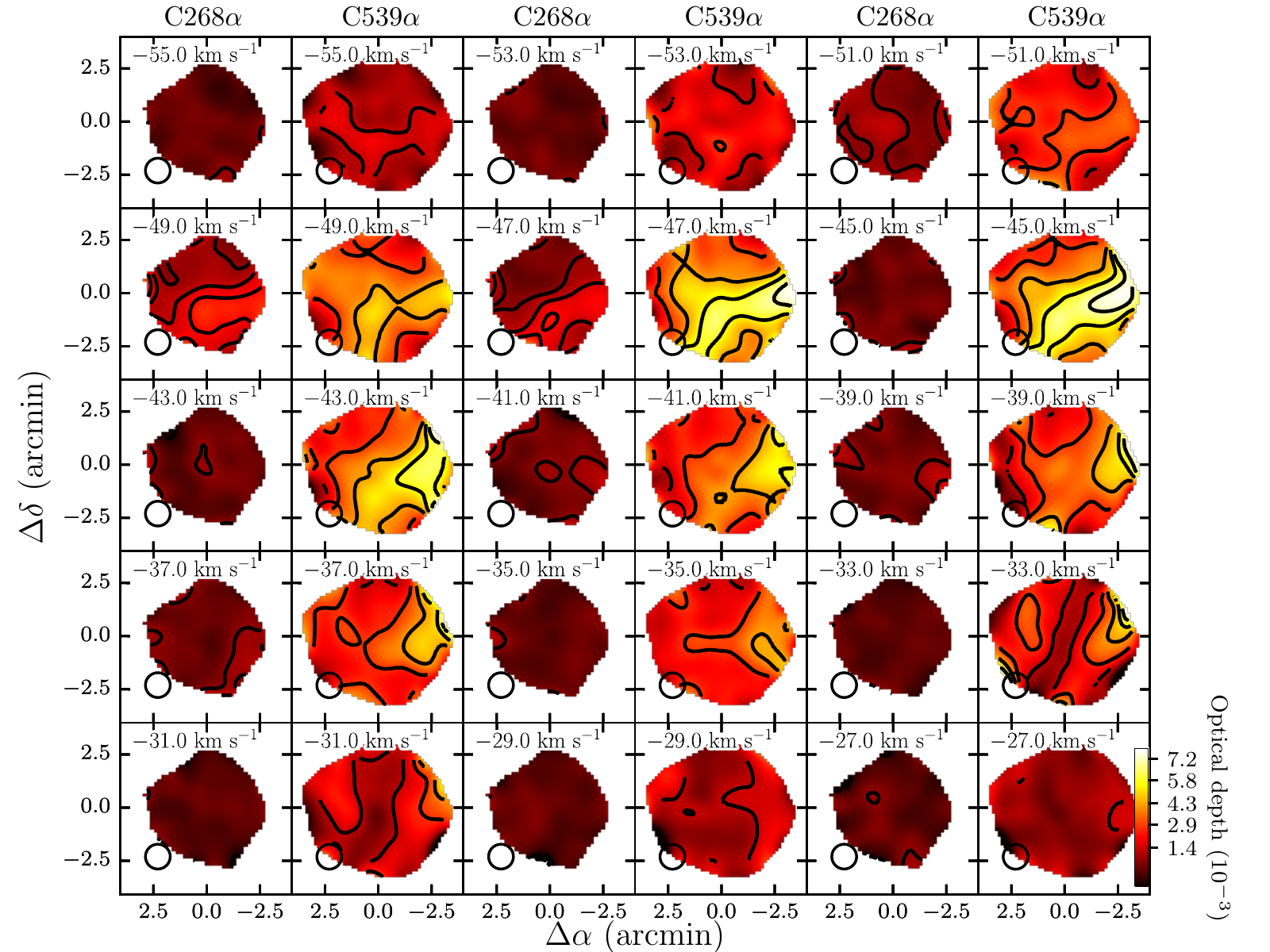}
 \end{center}
 \caption{C$268\alpha$ emission and C$539\alpha$ absorption channel maps.
 The spectral axis of the C$268\alpha$ cube was convolved to a $2.4$~km~s$^{-1}$ velocity resolution and reggrided to match the spectral axis of the C$539\alpha$ cube.
 The colour scale is the same for both C$268\alpha$ emission and C$539\alpha$ absorption.
 The contours start at $3\sigma$ with $\sigma=2.5\times10^{-4}$ for C$268\alpha$ and $\sigma=4.8\times10^{-4}$ for C$539\alpha$.
 The $70\arcsec$ resolution of the channel maps is shown in the bottom left corner of each map.}
 \label{fig:chlba}
\end{figure*}

Here we present spatially resolved C$268\alpha$ and C$539\alpha$ optical depth cubes. 
The C$357\alpha$ and C$494\alpha$ maps will be shown later (in \S~\ref{ssec:moms}), as these show a spatial distribution very similar to that of the C$268\alpha$ and C$539\alpha$ lines (Figures~\ref{fig:chwsrt} and \ref{fig:chlba}).

C$268\alpha$ channel maps at velocities around that of the $-47$~km~s$^{-1}$ velocity component are shown in Figure~\ref{fig:chwsrt}.
These maps show that the gas is predominantly concentrated to the southwest of Cas~A.
At around $-47$~km~s$^{-1}$ there is emission in an elongated structure running from Cas~A's western hotspot to its south.
This has been labelled with a white line between the W and S (Figure~\ref{fig:chwsrt}).
Higher resolution OH observations show that there are three OH clumps over this W-S structure at a velocity $\sim-47$~km~s$^{-1}$ \mbox{\citep[clumps B, D and E,][]{Bieging1986}}.
However, since OH and CRRLs do not trace exactly the same gas, we cannot use this as evidence that the W-S structure is a collection of clumps.
With the resolution of the cubes presented in this work it is not possible to distinguish if this is a filament or unresolved clumps.

Channel maps showing velocities corresponding to the Perseus arm features of the C$268\alpha$ and C$539\alpha$ lines are presented in Figure~\ref{fig:chlba}.
Here we use the velocity averaged C$268\alpha$ cube to emphasise features close to $-38$~km~s$^{-1}$.
Emission from the C$268\alpha$ $-38$~km~s$^{-1}$ velocity component is located in the western side of Cas~A as well as in the northeast, with a clump close to the centre of Cas~A between $-43$ and $-41$~km~s$^{-1}$.
This central clump is also identified in OH and CO \mbox{\citep{Bieging1986,Wilson1993}}.

In the C$539\alpha$ maps we can see similar structures to those seen in the C$268\alpha$ maps, albeit with more detail due to the higher signal-to-noise ratio of the LOFAR data.
The W-S structure is visible between $-49$ and $-37$~km~s$^{-1}$.
The larger extent in velocity is partially due to (radiation or pressure) broadening of the lines at high $n$ \citep[e.g.,][]{Salgado2017b}.
The spatial distribution of the C$539\alpha$ $-38$~km~s$^{-1}$ velocity component is harder to interpret due to the blending of the Perseus arm velocity components.
If we assume that absorption at the more positive velocities is mainly due to the $-38$~km~s$^{-1}$ velocity component, then the absorption close to $-33$~km~s$^{-1}$ should be representative.
At this velocity we see that the absorption comes primarily from the east and west of Cas~A, like the C$268\alpha$ emission from this velocity component.

\subsection{CRRL properties}
\label{ssec:moms}

The properties of low frequency CRRLs can be used to determine the gas electron density, temperature and pressure.
These properties imprint their signature in the line integrated optical depth as a function of principal quantum number $n$ \mbox{\citep[e.g.][]{Shaver1975,Salgado2017a}}.
Additionally, the change in line width with $n$, caused by radiation and pressure broadening, also provides information about the gas properties \mbox{\citep[e.g.][]{Shaver1975,Salgado2017a}}.

To determine the line properties over the face of Cas~A on a pixel-by-pixel basis we fit the line profiles and construct moment maps.
In order to fit the line profiles we first determine where the lines are detected by using the moment masking method as refined by \citet{Dame2011}.
In this method the data cube is smoothed to a resolution which is two times the original cube resolution in the spatial and spectral directions.
Using the smoothed cube we search for significant detections by requiring that the signal-to-noise ratio is above some threshold level.
For each pixel/channel which shows a significant detection we also set the neighbouring pixels which are inside the convolution kernel as detections.
Then we fit the line profile only in those regions where there are significant detections.

\begin{figure*}
 \begin{center}
  \includegraphics[width=\textwidth]{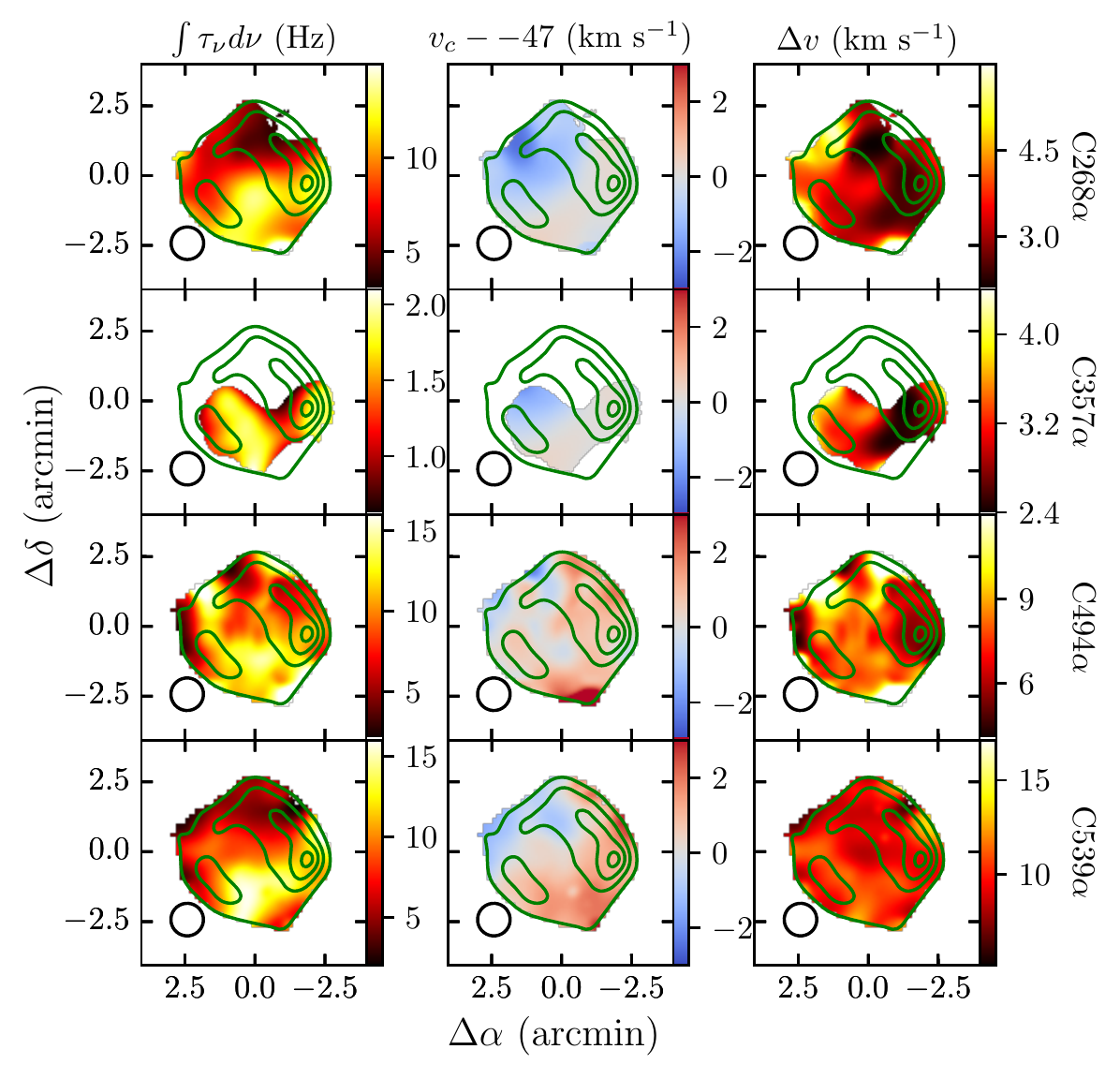}
 \end{center}
 \caption{Moment maps for the CRRLs at $-47$~km~s$^{-1}$.
 The {\it top row} shows the moment maps for the C$268\alpha$ line, the {\it upper middle row} for the C$357\alpha$ line, the {\it lower middle row} for the C$494\alpha$ line and the {\it bottom row} for the C$539\alpha$ line.
 The {\it left column} shows the integrated optical depth, the {\it middle column} the velocity centroid of the line with respect to $-47$~km~s$^{-1}$ and the {\it right column} the full width at half maximum of the line.
 The {\it green contours} show the $345$~MHz continuum from Cas~A at $51\arcsec\times45\arcsec$ resolution.
 The $70\arcsec$ resolution of the moment maps is shown in the bottom left corner of each map.
 At the edges of Cas~A the signal-to-noise ratios are lower due to the fainter continuum.}
 \label{fig:mom-47}
\end{figure*}

To determine an optimum threshold level we tested using synthetic data cubes.
In this test we varied the threshold level between one and ten times the noise in the synthetic data cube and compared the recovered moment $0$ with the known input.
This test shows that if we use a threshold of three times the noise in the smoothed cube, then the line properties can be recovered with no significant deviation from the input data.
For these observations this means that for the higher signal-to-noise line at $-47$~km~s$^{-1}$ we should recover most of the line structure. 
However, for the weaker velocity component at $-38$~km~s$^{-1}$, we are likely to recover only the brightest regions.

To fit the RRLs with principal quantum number $268$ we use Gaussian line profiles.
We fit up to three CRRLs close to $-47$~km~s$^{-1}$, $-38$~km~s$^{-1}$ and $0$~km~s$^{-1}$.
Once we have the line properties from the $n=268$ CRRLs, we use their second and third moments to guide the fit for the higher $n$ lines.
In the case of the $n=357$ lines this is necessary given the lower signal-to-noise ratio.
For the $n=494$ and $539$ lines this is done to guide the separation of the blended line profiles.
This relies on the assumption that the CRRLs at different frequencies will trace gas with similar properties.
Studies which cover a larger frequency range than the one studied here show that the line properties can be accurately modelled by a single set of gas properties \citep[][]{Oonk2017}.
With this the line centroid should be the same for different $n$ lines, and for $n\leq500$ the line profile is dominated by the gas thermal motion \citep{Salas2017}, which does not depend on frequency.

To fit the C$357\alpha$ lines we use two Gaussian profiles, one for the $-47$~km~s$^{-1}$ velocity component and one for $-38$~km~s$^{-1}$.
When fitting we fix the line centroid and line width to those of the C$268\alpha$ line.
This leaves the line amplitude as the only free parameter.

\begin{figure*}
 \begin{center}
  \includegraphics[width=\textwidth]{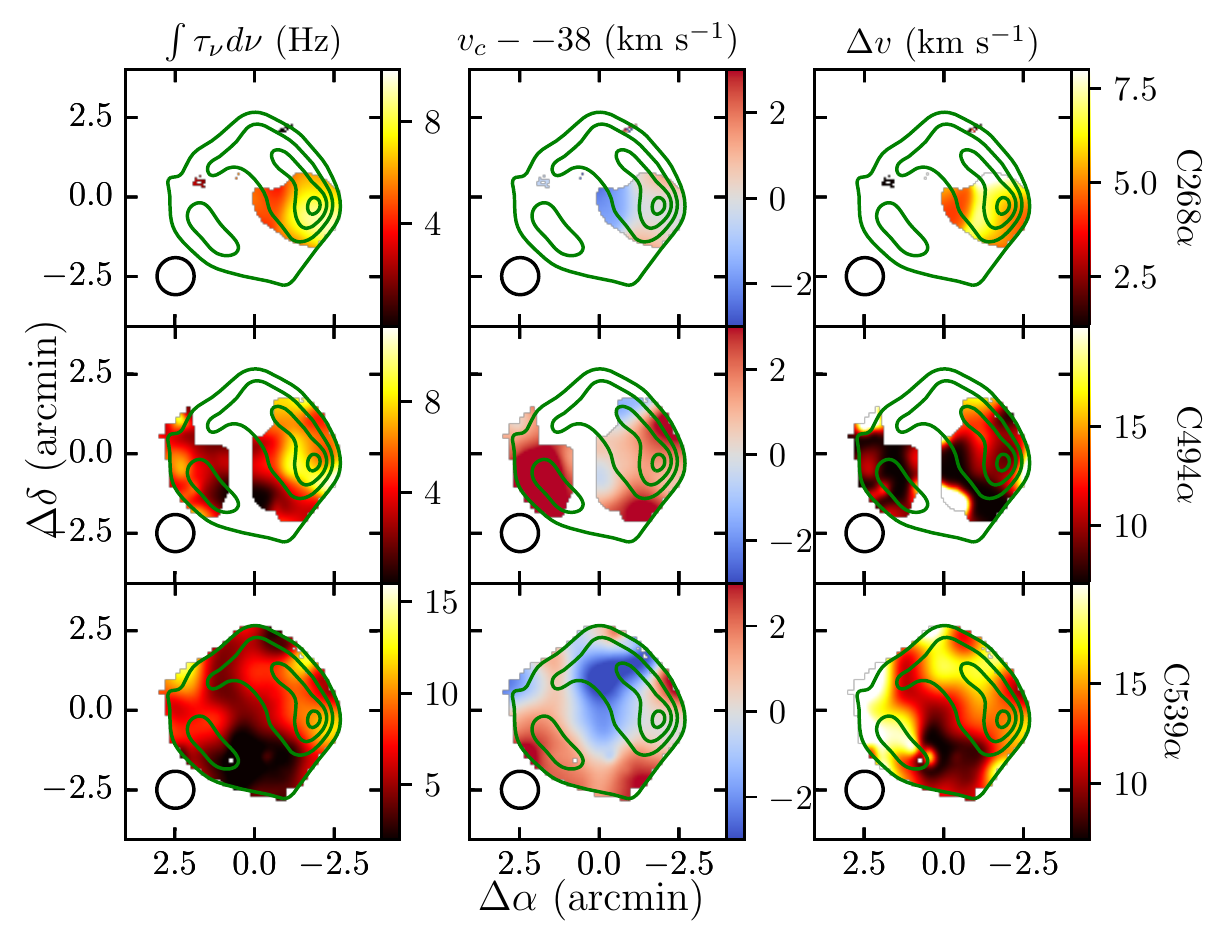}
 \end{center}
 \caption{Moment maps for the CRRLs at $-38$~km~s$^{-1}$.
 The {\it top row} shows the moment maps for the C$268\alpha$ line, the {\it middle row} for the C$494\alpha$ line and the {\it bottom row} for the C$539\alpha$ line.
 The {\it left column} shows the integrated optical depth, the {\it middle column} the velocity centroid of the line with respect to $-38$~km~s$^{-1}$ and the {\it right column} the full width at half maximum of the line.
 Here we do not show the C$357\alpha$ line moments as they are similar to those of the C$268\alpha$ line.
 The {\it green contours} show the $345$~MHz continuum from Cas~A at $51\arcsec\times45\arcsec$ resolution.
 The $70\arcsec$ resolution of the moment maps is shown in the bottom left corner of each map.
 At the edges of Cas~A the signal-to-noise ratios are lower due to the fainter continuum.}
 \label{fig:mom-38}
\end{figure*}

To fit the C$494\alpha$ and C$539\alpha$ lines we use two Voigt profiles, one for the $-47$~km~s$^{-1}$ velocity component and one for $-38$~km~s$^{-1}$.
When fitting we fix the line centroid and the Doppler core of the line profiles to those of the C$268\alpha$ lines.
This leaves the amplitude and Lorentz width as free parameters.
When there is no significant C$268\alpha$ line emission we adopt the median values from the C$268\alpha$ moments as initial guesses for the line parameters, but we allow them to vary.
This is mostly the case for the line at $-38$~km~s$^{-1}$.

The moment maps for the $n=268$ RRL at $-47$~km~s$^{-1}$ are shown in the top row of Figure~\ref{fig:mom-47}.
Here we see that emission from the $-47$~km~s$^{-1}$ velocity component extends almost all over the face of Cas~A, with a lower integrated optical depth to the north of the remnant.
The moment $0$, or velocity integrated optical depth, map for the $-47$~km~s$^{-1}$ velocity component shows that most of the emission comes from the W-S structure.
The moment $1$, or optical depth weighted velocity centroid, map shows that over the W-S structure the velocity remains constant. 
The moment $2$, or full width at half maximum, map shows that the line is narrow in the West and broadens to the East of Cas~A.
Towards the North-east of Cas~A the $-47$~km~s$^{-1}$ C$268\alpha$ line is broader by a factor of $\sim2$ with respect to the W-S structure.
The CRRL at $0$~km~s$^{-1}$ is not displayed because at $70\arcsec$ resolution the signal-to-noise ratio is lower than three in individual pixels.

The moment maps for the C$357\alpha$ $-47$~km~s$^{-1}$ velocity component are shown in the middle panels of Figure~\ref{fig:mom-47}.
These show that the velocity integrated optical depth of the $-47$~km~s$^{-1}$ velocity component is larger in the W-S structure, similar to that observed in the C$268\alpha$ line.

The moment maps for the C$539\alpha$ line at $-47$~km~s$^{-1}$ are shown in the bottom panels of Figure~\ref{fig:mom-47}.
These show that most of the C$539\alpha$ absorption from the $-47$~km~s$^{-1}$ velocity component comes from the W-S structure, in accordance with the lower $n$ lines.
The moment $2$ maps suggests that the $-47$~km~s$^{-1}$ velocity component is broader towards the south-east of Cas~A, however the line width is consistent with a constant value over the face of Cas~A.

The moment maps for the C$268\alpha$ and C$539\alpha$ $-38$~km~s$^{-1}$ velocity component are shown in Figure~\ref{fig:mom-38}.
The C$268\alpha$ line is only detected towards the western hotspot of Cas~A.
In contrast, the C$539\alpha$ line is detected almost all over the face of Cas~A. 
Both maps show that the peak integrated optical depth is located towards the western hotspot of Cas~A.
For the C$357\alpha$ line at $-38$~km~s$^{-1}$ the spatial distribution is similar to that of the C$268\alpha$ at the same velocity, a patch $\sim0.4\arcmin$ towards the south of the western hotspot of Cas~A.

The moment $2$ maps for the $-38$~km~s$^{-1}$ component shows that the line is broader towards the northern half of Cas~A, with the broadest line towards the east.
The minimum line broadening is observed towards the centre and south of Cas~A. 
This resembles the ridge of CRRL absorption that passes through the centre of Cas~A between $-33$ and $-31$~km~s$^{-1}$ (Figure~\ref{fig:chlba}).

\begin{figure*}
 \begin{center}
  \includegraphics[width=\textwidth]{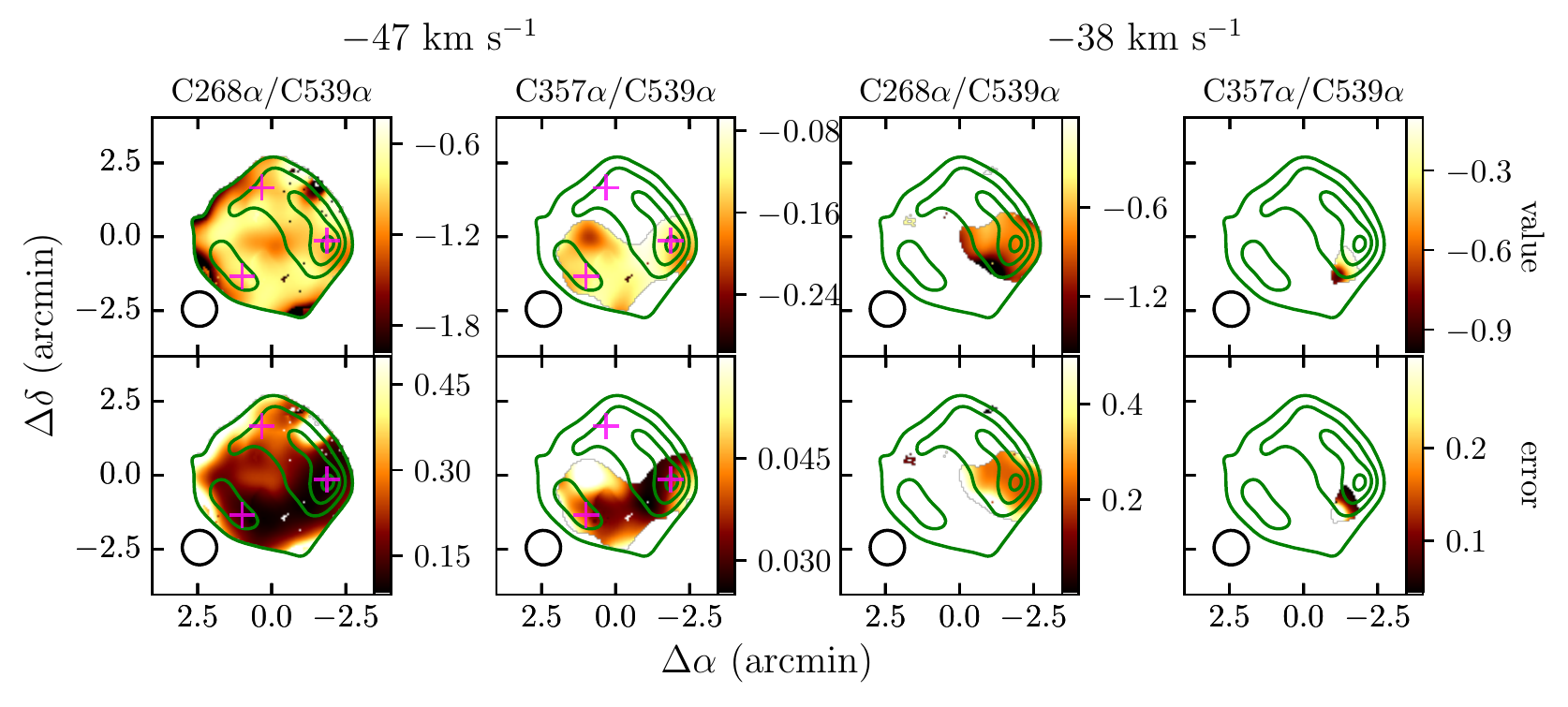}
 \end{center}
 \caption{Ratios between different CRRLs at different velocities.
 The ratios shown are the $R^{268}_{539}=$C$268\alpha$/C$539\alpha$ and $R^{357}_{539}=$C$357\alpha$/C$539\alpha$ line ratios.
 The {\it top panels} show the ratio value and the {\it bottom panels} the error on the ratio.
 The {\it text on top of each column} indicates which ratio is shown in the corresponding column, and a {\it text label} on the top of the Figure shows the velocity component.
 The ratios show only regions where both lines involved have been detected with a signal-to-noise ratio larger than three.
 The {\it pink crosses} in the top panel show the regions used in Figure~\ref{fig:crrlmod}.
 The $70\arcsec$ resolution of these maps is shown in the bottom left corner of each map.}
 \label{fig:crrlratios}
\end{figure*}

\subsection{Physical conditions from CRRLs}
\label{ssec:gaspropscrrls}

The change in the CRRL profile as a function of principal quantum number has the signature of the gas physical conditions imprinted on it \mbox{\citep{Shaver1975,Salgado2017a}}.
The gas properties which can be determined using CRRLs are its electron density $n_{\rm{e}}$, electron temperature $T_{\rm{e}}$ and the intensity at $100$~MHz of the radiation field the carbon atoms are immersed in, $T_{r,100}$.

To determine the CRRL properties we use the models of \mbox{\citet{Salgado2017a}}.
To solve the level population problem we assume that collisions with atomic hydrogen and electrons set the relative population of carbon ions in the $^{2}P_{1/2}$ and $^{2}P_{3/2}$ states.
For the collisional rates we adopt the values of \mbox{\citet{Tielens1985a}} for collisions with hydrogen and those of \mbox{\citet{Hayes1984}} for collisions with electrons.
We adopt electron and carbon abundances of $3\times10^{-4}$ while solving the level population problem.
However, when converting to per hydrogen atom quantities, we adopt electron and carbon abundances of $1.5\times10^{-4}$.
The effect of these assumptions is small ($\sim10\%$), and will be discussed later.
For the $\ell$-changing collisional rates we use the semi-classical formulation of \citet{Vrinceanu2012} incorporated into the \citet{Salgado2017a} models.
Additionally, when solving the level population problem, a radiation field with a power law shape is included.
This radiation field has a spectral index of $-2.6$, similar to the observed spectral index of Galactic synchrotron emission \citep[e.g.,][]{Oliveira-Costa2008,Zheng2016}, and its intensity is defined at $100$~MHz by $T_{r,100}$ \citep{Shaver1975,Salgado2017a}.

In order to model the gas properties we assume that the radiation field the gas is immersed in is constant.
Since we are studying gas on scales of $\lesssim1$~pc and the gas is at a distance of $\gtrsim220$~pc from Cas~A \citep[e.g.,][]{Kantharia1998,Salas2017}, the possible contribution of Cas~A to the radiation field \citep{Stepkin2007} will change by a negligible amount over the observed structure.
Additionally, there are no other known strong, low-frequency, discrete radiation sources in the field.
Considering this, it seems reasonable to assume that the gas is immersed in a constant radiation field.
Following the results of \citet{Oonk2017} we adopt $T_{r,100}=1400$~K.

Given the C$268\alpha$, C$357\alpha$, C$494\alpha$ and C$539\alpha$ velocity integrated optical depths we explore how these can be used to constrain the gas properties.
We do not attempt to model the change in integrated optical depth as a function of $n$, like \citet{Oonk2017} did, given that the number of free parameters is similar to the number of data points.
Instead, we use the ratios between these lines.
We will use the notation $R^{n}_{n^{\prime}}$ to denominate the C$n\alpha$/C$n^{\prime}\alpha$ line ratio, e.g., C$357\alpha$/C$539\alpha$ line ratio will be $R^{357}_{539}$.

To generate $R^{268}_{539}$, $R^{268}_{494}$ and $R^{357}_{539}$ from the observations we use the respective velocity integrated optical depth maps (Figures~\ref{fig:mom-47} and \ref{fig:mom-38}).
These line ratios are shown in Figure~\ref{fig:crrlratios}.
In the case that one of the lines is not detected in one of the pixels we adopt the $3\sigma$ upper/lower limit on the integrated optical depth for the non detection.
If two lines are not detected on a pixel we do not attempt to constrain the gas physical conditions.
This limits our analysis for the $-38$~km~s$^{-1}$ velocity component to the western hotspot of Cas~A, where the C$268\alpha$ line is detected (Figure~\ref{fig:mom-38}).

\begin{figure*}
 \begin{center}
  \includegraphics[width=\textwidth]{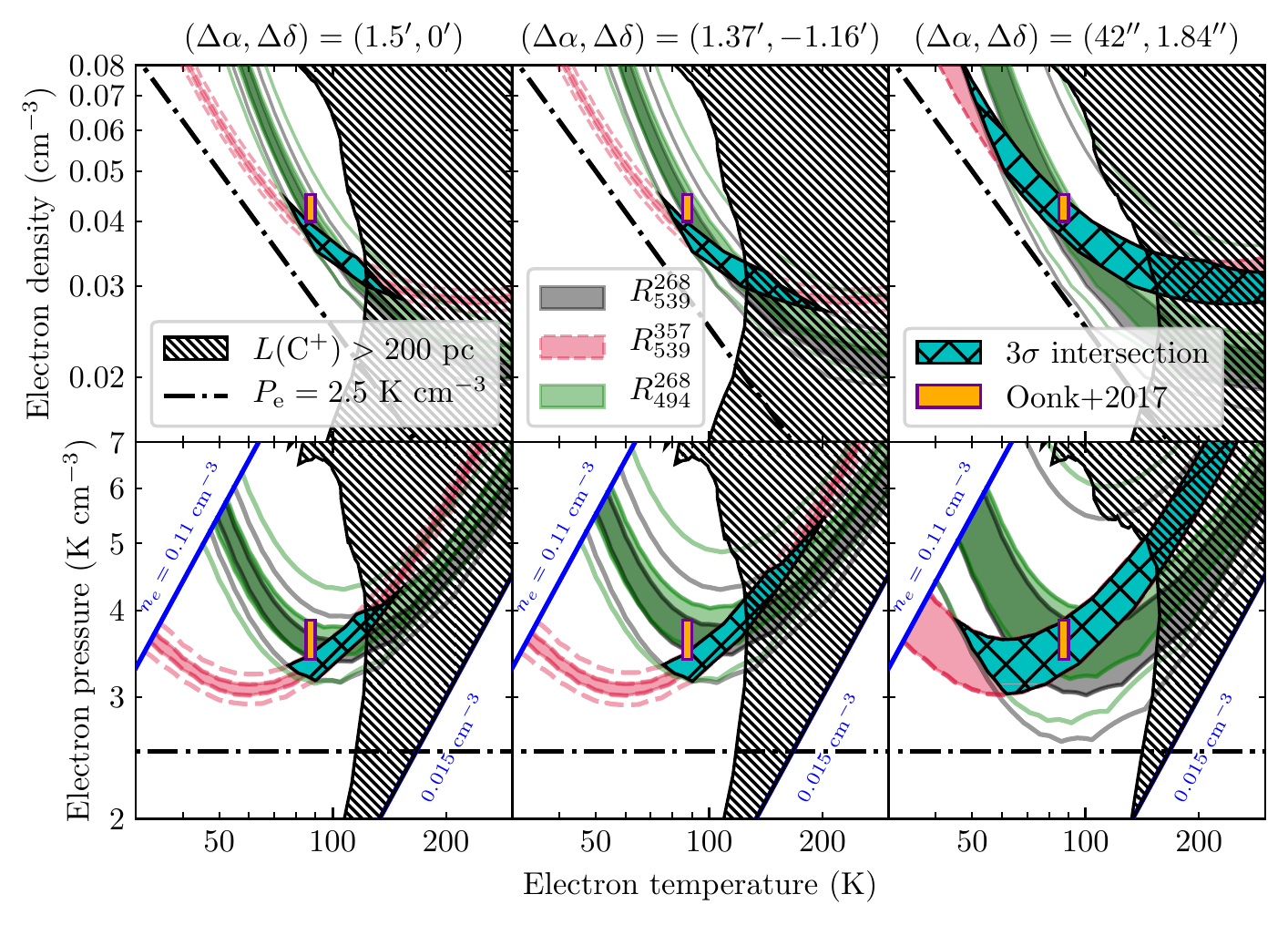}
 \end{center}
 \caption{Constraints imposed by the $R^{268}_{539}=$C$268\alpha$/C$539\alpha$, $R^{357}_{539}=$C$357\alpha$/C$539\alpha$ and $R^{268}_{494}=$C$268\alpha$/C$494\alpha$ line ratios in the $n_{\rm{e}}\mbox{--}T_{\rm{e}}$ plane ({\it top panels}) and in the $P_{\rm{e}}\mbox{--}T_{\rm{e}}$ plane ({\it bottom panels}).
 The constraints are shown for the $-47$~km~s$^{-1}$ velocity component at three different positions in the map.
 The positions, relative to the map centre, are shown on the top of the corresponding column.
 These positions are shown in Figures~\ref{fig:crrlratios} and \ref{fig:crrlgas} as green crosses.
 The {\it orange rectangle with purple borders} shows the gas physical conditions derived by \citet{Oonk2017}.
 The {\it dot-dashed line} shows a curve of constant electronic pressure.
 The {\it cyan region hatched with black lines} shows the intersection of the constraints imposed by the $R^{268}_{539}$, $R^{357}_{539}$ and $R^{268}_{494}$ line ratios if we consider their $3\sigma$ ranges.
 The {\it densely hatched region} shows where the required ionized carbon path length is larger than $200$~pc.
 {\it Blue lines} in the bottom row show the limits of the grid of CRRL models.}
 \label{fig:crrlmod}
\end{figure*}

\begin{figure}
 \begin{center}
  \includegraphics[width=0.49\textwidth]{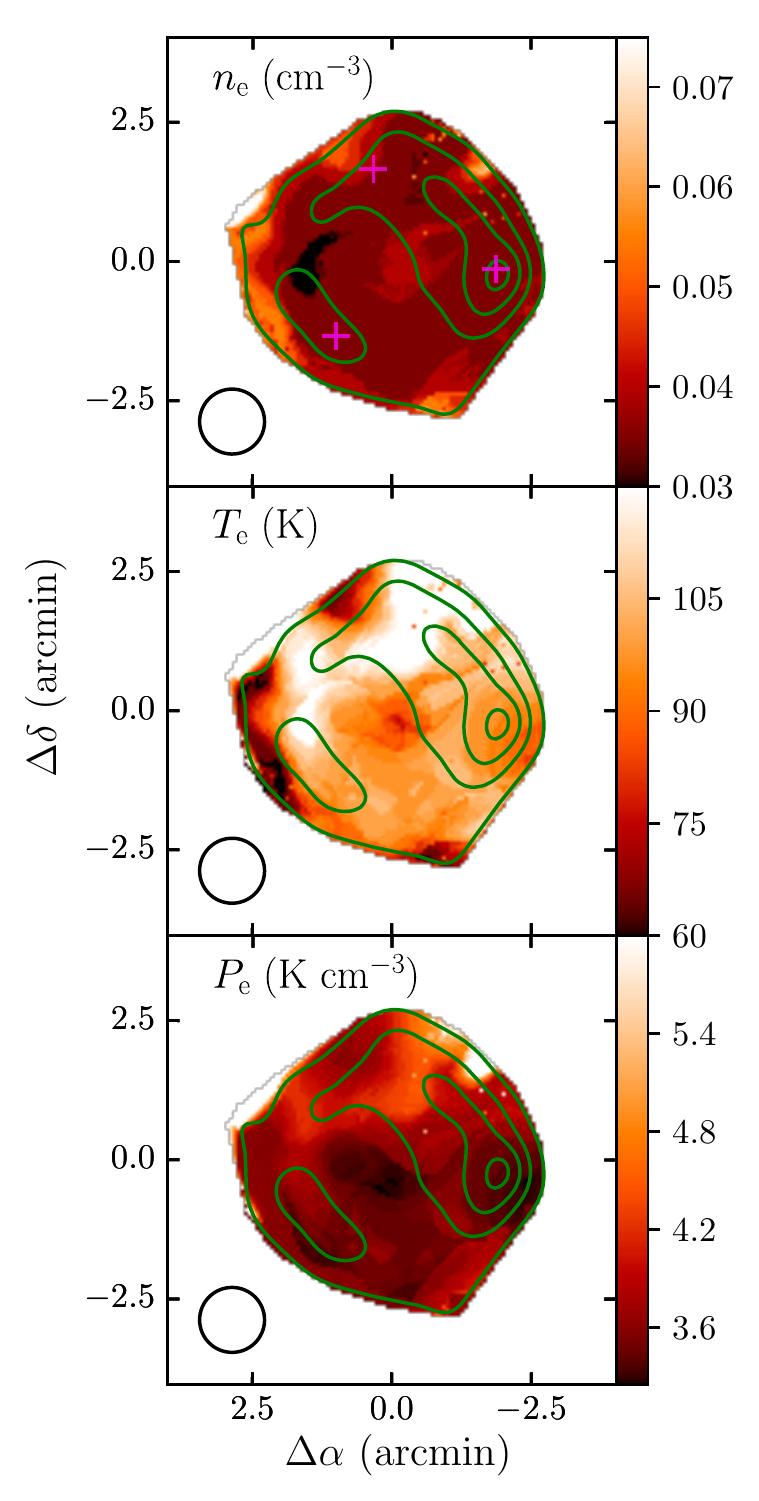}
 \end{center}
 \caption{Gas properties derived from the CRRL ratios for the Perseus arm component at $-47$~km~s$^{-1}$.
 The {\it top panel} shows the electron density, the {\it middle panel} the electron temperature and the {\it bottom panel} the electron pressure.
 These are derived under the assumption of a constant radiation field of $1400$~K at $100$~MHz.
 The {\it pink crosses} in the top panel show the regions used in Figure~\ref{fig:crrlmod}.
 The $70\arcsec$ resolution of these maps is shown in the bottom left corner of each map.}
 \label{fig:crrlgas}
\end{figure}

The line ratios as a function of gas properties in the $n_{\rm{e}}\mbox{--}T_{\rm{e}}$ plane are shown in Figure~\ref{fig:crrlmod} for three different locations in the map (see Figures~\ref{fig:crrlratios} and \ref{fig:crrlgas}).
As a result of the C$268\alpha$ line being in emission and the C$539\alpha$ line in absorption $R^{268}_{539}$ produces contours which have a similar shape to the constraint imposed by the transition from emission to absorption \citep{Salgado2017b}.
The situation is similar for $R^{268}_{494}$.
However intersecting the $R^{357}_{539}$ constraint with either $R^{268}_{539}$ or $R^{268}_{494}$ restricts the range of allowed $n_{\rm{e}}\mbox{ and }T_{\rm{e}}$ values.
The left panel of Figure~\ref{fig:crrlmod} shows that the CRRL ratios used can constrain the gas properties in regions with high signal-to-noise ratio detections.
Yet it also reveals that when one of the lines is not detected it is not possible to constrain the gas properties (e.g., right panel in Figure~\ref{fig:crrlmod}).
In this case the line ratios constrain the gas electron density, and place a lower limit on the gas temperature.
An upper limit on the gas temperature can be obtained if we consider the implied gas path length.
If we restrict the models to path lengths smaller than $200$~pc, this effectively puts an upper limit on the electron temperature of $\approx160$~K (Figure~\ref{fig:crrlmod}).
We adopt an upper limit of $200$~pc since we do not expect the gas structures to be larger than this in the line of sight direction.

\begin{table}
\begin{center}
\caption{Grid of CRRL models}
\begin{tabular}{lccc}
\hline
\hline
Parameter & Notation & Value & Step \\
\hline
Electron density & $n_{\rm{e}}$ & $0.01$--$0.1$~cm$^{-3}$ & $0.005$~cm$^{-3}$ \\
Electron temperature & $T_{\rm{e}}$ & $10$--$400$~K & $5$~K \\
Radiation field      & \multirow{2}{*}{$T_{\rm{r,100}}$} & $800$--$2000$~K & $400$~K \\
at $100$~MHz         &                                   & \multicolumn{2}{c}{and $1400$~K}  \\
\hline
\end{tabular}
\label{tab:crrlgrid}
\end{center}
\end{table} 

Maps with the electron density, temperature and pressure constraints derived from the line ratio analysis for the $-47$~km~s$^{-1}$ velocity component are shown in Figure~\ref{fig:crrlgas}.
The electron density shows almost no variation over the face of Cas~A, while the electron temperature and pressure show a slight decrease towards the South.
The electron density has an almost constant value over the face of Cas~A, however this largely due to the resolution of the model grid.
In the density axis we have $18$ resolution elements, while on the temperature axis we have $78$ (Table~\ref{tab:crrlgrid}).
The discrete nature of the grid of models used to determine the electron temperature and density also results in abrupt changes in the gas properties.
Additionally, this discreteness can produce patches which have sizes smaller than the spatial resolution of the data.
This effect is particularly notorious in the $T_{\rm{e}}$ map (Figure~\ref{fig:crrlgas}).

As it is evident in Figure~\ref{fig:crrlmod}, we are constraining the gas properties to a given range.
The size of this range will depend on the error bars of each pixel.
For pixels with high signal-to-noise (left and centre panels in Figure~\ref{fig:crrlmod}) the uncertainty in electron density is about $\sim30\%$ and in electron temperature $\sim25\%$.
While on pixels with lower signal to noise (right panel in Figure~\ref{fig:crrlmod}), the uncertainty can be of a factor of three or more.
A change of about $25\%$ in electron temperature translates into a $65\%$ change in emission measure.
In Table~\ref{tab:gasprops} we present the gas properties averaged over the face of Cas~A.
In this Table we provide the parameter ranges if we consider the $1\sigma$ uncertainties in the observed line ratios.
In terms of the spatial distribution, this shows little change when we consider the uncertainties.
The biggest change is on the mean value.

To estimate the hydrogen density from the electron density we assume that $94\%$ and $100\%$ of free electrons come from carbon for the $-47$ and $-38$~km~s$^{-1}$ velocity components respectively \citep{Oonk2017}.
Additionally, we adopt a carbon abundance relative to hydrogen of $1.5\times10^{-4}$ \citep{Sofia1997}.
With this, the hydrogen density is $210$--$360$ and $200$--$470$~cm$^{-3}$ for the $-47$ and $-38$~km~s$^{-1}$ velocity components respectively.

The largest uncertainty in the derived gas properties comes from the separation between the $-47$ and $-38$~km~s$^{-1}$ velocity components for the $n>400$ lines.
We test this effect by varying the integrated optical depth of the C$539\alpha$ line and comparing the derived values of $n_{\rm{e}}$ and $T_{\rm{e}}$.
We find that the electron temperature has an almost linear relation with the integrated optical depth of the C$539\alpha$ line, i.e. if we decrease the C$539\alpha$ line integrated optical depth by $30\%$, then the derived electron temperature decreases by $30\%$.
For the electron density the change is less pronounced.
A change of $30\%$ in the integrated optical depth of the C$539\alpha$ line results in a change of less than $15\%$ in the electron density.

\begin{figure*}
 \begin{center}
  \includegraphics[width=\textwidth]{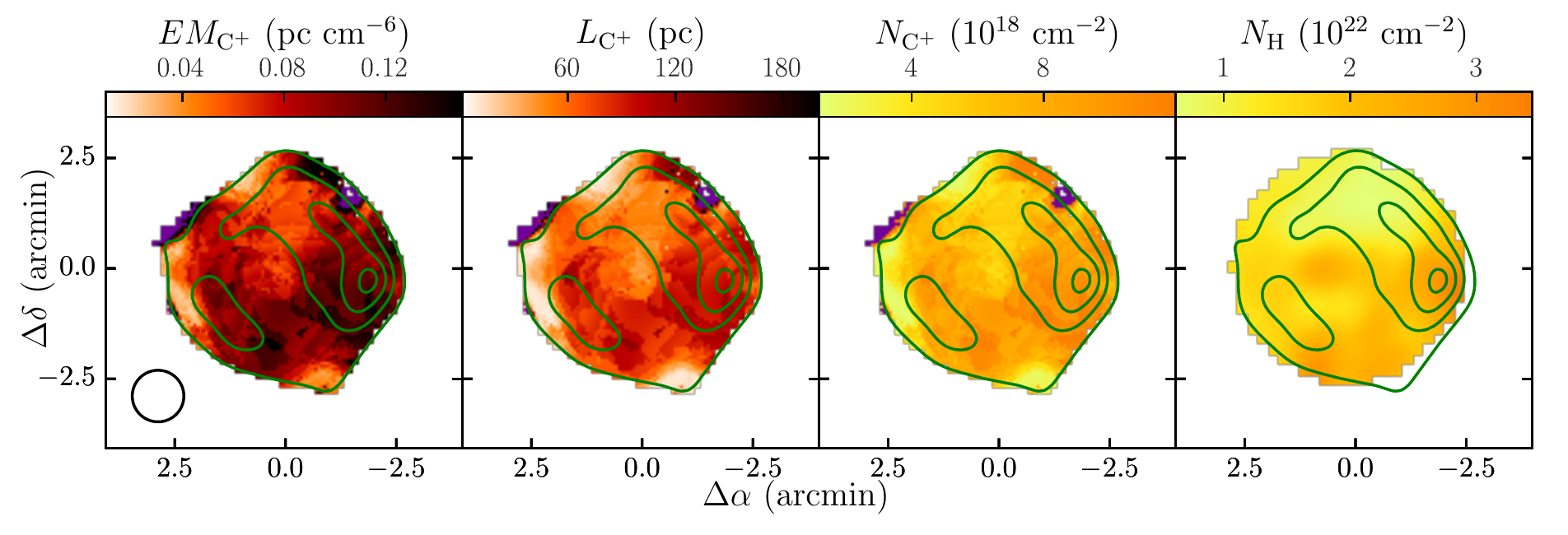}
 \end{center}
 \caption{C$^{+}$ emission measure $EM_{\rm{CII}}$, path length $L_{\rm{CII}}$ and column density of ionized carbon $N(\rm{CII})$ derived from the CRRL analysis for the Perseus arm component at $-47$~km~s$^{-1}$, and the hydrogen column density map derived from the dust analysis by \citet{deLooze2017}. 
 The {\it leftmost column} shows the emission measure, the {\it middle left column} the path length, the {\it middle right column} the column density of ionized carbon and the {\it rightmost column} the hydrogen column density derived from analysis of the dust emission towards Cas~A \citep{deLooze2017}.
 Regions where we cannot constrain the gas properties given the parameter space explored and observational uncertainties are shown with a colour not present in the colour bar.
 The $70\arcsec$ resolution of these maps is shown in the bottom left corner of each map.
 The {\it green contours} show the $345$~MHz continuum from Cas~A at $51\arcsec\times45\arcsec$ resolution.}
 \label{fig:crrlemelnc}
\end{figure*}

Using the electron density and temperature maps and the C$268\alpha$ integrated optical depth we compute the ionized carbon emission measure, $EM_{\rm{CII}}$, column density, $N(\rm{CII})$, and its path length along the line of sight, $L_{\rm{CII}}$.
To determine the column density and $L_{\rm{CII}}$ we assume that $94\%$ of the free electrons come from ionized carbon \citep{Oonk2017}.
Maps with $EM_{\rm{CII}}$, $N(\rm{CII})$ and $L_{\rm{CII}}$ for the $-47$~km~s$^{-1}$ velocity component are shown in Figure~\ref{fig:crrlemelnc}.
These show a similar structure to the C$268\alpha$ integrated optical depth map, with larger values towards the South and West of Cas~A.
The structures smaller than the beam size are due to the strong dependence of the integrated optical depth on the electron temperature ($\propto T_{\rm{e}}^{-5/2}$), and the discrete nature of the model grid used.
For a constant electron and carbon density the only variation in the emission measure comes from variations in the path length of the gas along the line of sight.
This will also be reflected in the column density $N(\rm{CII})=L_{\rm{CII}}n_{\rm{CII}}$.

\begin{table}
\begin{center}
\begin{threeparttable}
\caption{Ranges of the mean gas properties over the face of Cas~A}
\begin{tabular}{lcc}
\hline
\hline
Gas property & \multicolumn{2}{c}{Velocity component} \\
             & $-47$ km~s$^{-1}$      & $-38$ km~s$^{-1}$\\
\hline
$T_{\rm{e}}$ (K)           & $71$--$137$       & $50$--$145$  \\
$n_{\rm{e}}$ (cm$^{-3}$)   & $0.03$--$0.055$   & $0.03$--$0.07$ \\
$P_{\rm{e}}$ (K cm$^{-3}$) & $3.1$--$4.95$     & $2.7$--$4.8$ \\
\hline
$EM_{\rm{CII}}$ (pc~cm$^{-6}$)       & $0.046$--$0.26$ & $0.02$--$0.21$ \\
$N_{\rm{CII}}$ ($10^{18}$ cm$^{-2}$) & $3.4$--$25.9$   & $1.6$--$18.1$  \\
$L_{\rm{CII}}$ (pc)                  & $27$--$182$     & $10$--$180$ \\
\hline
$n_{\rm{H}}$\tnote{a} (cm$^{-3}$)            & $210$--$360$    & $200$--$470$ \\
$N_{\rm{H}}$\tnote{a} ($10^{22}$ cm$^{-2}$)   & $2.3$--$17.3$   & $1$--$12$ \\
$A_{\rm{V}}$\tnote{b}                         & $11.5$--$360$    & $200$--$470$ \\
\hline
\end{tabular}
\label{tab:gasprops}
\begin{tablenotes}\footnotesize
\item[a] Assuming a carbon abundance relative to hydrogen of $1.5\times10^{-4}$ \citep{Sofia1997} and that $94\%$ and $100\%$ of the free electrons come from ionized carbon for the $-47$ and $-38$ km~s$^{-1}$ velocity components respectively \citep{Oonk2017}.
\item[b] Adopting $N_{\rm{H}}=(2.08\pm0.02)\times10^{21}A_{\rm{V}}$~cm$^{-2}$ \citep{Zhu2017}.
\end{tablenotes}
\end{threeparttable}
\end{center}
\end{table}

We compare the derived gas physical conditions against the spatially unresolved work of \citet{Oonk2017} towards the same background source to check for any differences.
We focus on their study as it was the first one which was able to simultaneously explain the line width and integrated optical depth change with $n$.
Additionally, this work and that of \citet{Oonk2017} use the same models, which reduces the need to account for different assumptions in the modelling.
If we compare Table~\ref{tab:gasprops} with Table~7 of \citet{Oonk2017} we see that our results, averaged over the face of Cas~A, are consistent.

\subsection{$158$~$\mu\rm{m}$-\rm{[CII]} {\bf line properties}}

The $158$~$\mu$m-[CII] line is the main coolant of the neutral diffuse ISM \citep[e.g.][]{Wolfire1995,Wolfire2003}.
Here we present the properties of the spatially resolved, velocity unresolved, $158$~$\mu$m [CII] line.

\begin{figure}
 \begin{center}
  \includegraphics[width=0.49\textwidth]{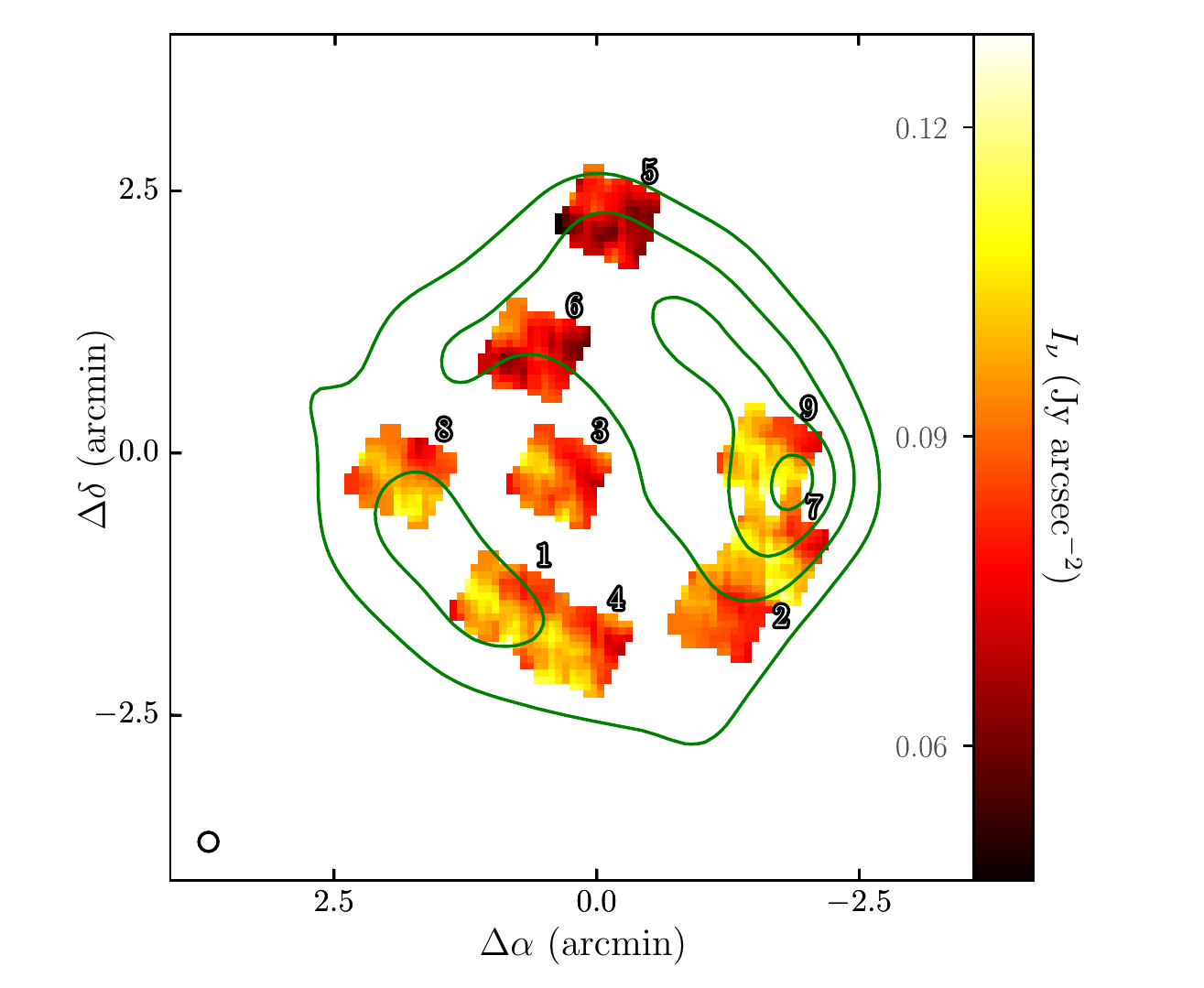}
 \end{center}
 \caption{Map of $158$~$\mu$m-[CII] line emission obtained with Herschel PACS.
 In these the [CII] line is unresolved in velocity, and only $\sim20\%$ of the surface of Cas~A is covered.
 This map is shown at its native resolution of $12\arcsec$.
 }
 \label{fig:ciimap}
\end{figure}

To determine the properties of the $158$~$\mu$m-[CII] line we fit a Gaussian profile to each of the nine PACS footprints.
We only fit one Gaussian component since the line is unresolved in velocity.
The best fit parameters of the Gaussian profile are presented in Table~\ref{tab:ciipars}.
These show little variation in the line frequency integrated intensity, but we do note that the lowest values are found in the northern footprints (5 and 6, see Figure~\ref{fig:ciimap}).
This could be due to the lower column densities found towards the north of Cas~A (see Figure~\ref{fig:crrlemelnc}).

\begin{table}
\begin{center}
\begin{threeparttable}
\caption{$158$~$\mu$m-[CII] Line properties}
\begin{tabular}{cccc}
\hline
\hline
Region & $v_{\rm{c}}$ & $\int I_{\nu}d\nu$ & $L_{\rm{[CII]}}$\\
 & (km s$^{-1}$) & ($10^{-5}$ erg cm$^{-2}$ s$^{-1}$ sr$^{-1}$) & (L$_{\odot}$)\\
\hline
1 & $-19\pm15$ & $6.70\pm0.11$ & $1.14\pm0.02$ \\ 
2 & $-20\pm15$ & $5.99\pm0.11$ & $1.02\pm0.02$ \\ 
3 & $-18\pm15$ & $6.15\pm0.09$ & $1.05\pm0.01$ \\ 
4 & $-20\pm15$ & $6.76\pm0.12$ & $1.15\pm0.02$ \\ 
5 & $-19\pm15$ & $4.65\pm0.22$ & $0.79\pm0.04$ \\ 
6 & $-23\pm15$ & $5.10\pm0.19$ & $0.87\pm0.03$ \\ 
7 & $-19\pm15$ & $6.92\pm0.08$ & $1.18\pm0.01$ \\ 
8 & $-19\pm15$ & $6.74\pm0.15$ & $1.15\pm0.02$ \\ 
9 & $-15\pm15$ & $7.14\pm0.11$ & $1.22\pm0.02$ \\ 
\hline
\end{tabular}
\label{tab:ciipars}
\end{threeparttable}
\end{center}
\end{table}

If we take the observed luminosity of the $158$~$\mu$m-[CII] line and compare it to the CRRL derived gas column density we obtain values of the order of $(1.4\pm0.1)\times10^{-26}$~erg~s$^{-1}$~(H-atom)$^{-1}$.
This cooling [CII] rate is is somewhat less than the cooling rate derived from ultraviolet absorption line studies originating from the upper fine structure level in sighlines through nearby diffuse clouds \citep[$(3\mbox{--}10)\times10^{-26}$~erg~s$^{-1}$~(H-atom)$^{-1}$][]{Pottasch1979,Gry1992} but comparable to the average cooling rate of the Galaxy, $(2.65\pm0.15)\times10^{-26}$~erg~s$^{-1}$~(H-atom)$^{-1}$ \citep{Bennett1994}.
For the CRRL derived column densities (Table~\ref{tab:gasprops}), the $158$~$\mu$m-[CII] line will be optically thick \citep[e.g.,][]{Tielens1985a}.
If the [CII] line is optically thick, then the observed line does not account for the total line of sight ionized carbon column which in turn results in a lower [CII] cooling rate.

\section{Discussion}
\label{sec:discussion}

\subsection{Comparison with other tracers}

We compare the CRRL optical depth with lines which trace different components of the ISM.
These include diffuse atomic gas \citep[$21$~cm-HI,][]{Bieging1991}, diffuse molecular gas \citep[$18$~cm-OH,][]{Bieging1986}, translucent gas \citep[$492$~GHz--{[}CI{]},][]{Mookerjea2006} and dense molecular gas \mbox{\citep[CO,][]{Wilson1993,Liszt1999,Kilpatrick2014}}.

\subsubsection{Spatial distribution}

\begin{figure*}
 \begin{center}
  \includegraphics[width=\textwidth]{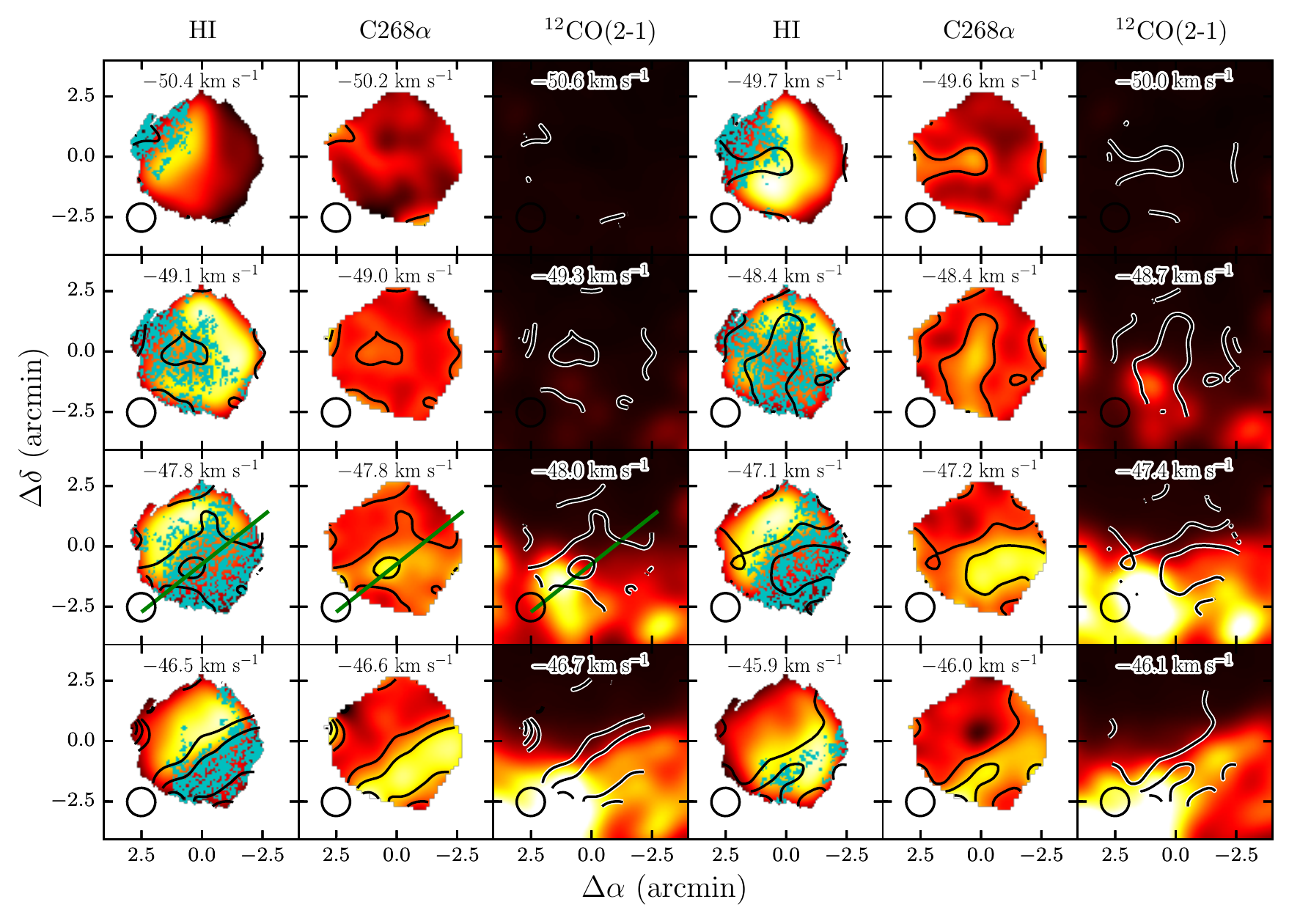}
 \end{center}
 \caption{Comparison between the optical depths of $21$~cm-HI, C$268\alpha$ and the $^{12}$CO$(2$--$1)$ line brightness.
 Masked pixels in the $21$~cm-HI optical depth maps are shown in cyan.
 The {\it green line} in the $-47.8$~km~s$^{-1}$ map shows the slice used to study the gas between the peaks of the $^{12}$CO$(2$--$1)$ and C$268\alpha$ lines.
 The $70\arcsec$ resolution of these maps is shown in the bottom left corner of each map.}
 \label{fig:hicrrlco}
\end{figure*}

A comparison between the optical depths of $21$~cm-HI, C$268\alpha$ and the $^{12}$CO$(2$--$1)$ line is presented in Figure~\ref{fig:hicrrlco}.
This shows that most of the C$268\alpha$ emission comes from regions where HI is saturated (cyan pixels in the HI maps).
The $^{12}$CO$(2$--$1)$ line also shows structures which are well correlated with the ones seen in C$268\alpha$ and $21$~cm-HI.
However, the peaks of CO emission are generally located outside the face of Cas~A, which does not allow for a direct comparison.
One exception is at a velocity of $-47.8$~km~s$^{-1}$, where a peak of $^{12}$CO$(2$--$1)$ emission is located over the face of Cas~A.
In this case the distance between the peaks of $^{12}$CO$(2$--$1)$ and C$268\alpha$ is $87\arcsec$.

The spatial distribution of $^{12}$CO$(2$--$1)$ shows that most of the gas at $-41$~km~s$^{-1}$ is located to the West of Cas~A, while the gas at $-36.5$~km~s$^{-1}$ extends from the west to the south east of Cas~A \citep{Kilpatrick2014}.
Both velocity components overlap towards the West of Cas~A.
This makes the distinction of these velocity components more difficult in this region.

\begin{figure}
 \begin{center}
  \includegraphics[width=0.49\textwidth]{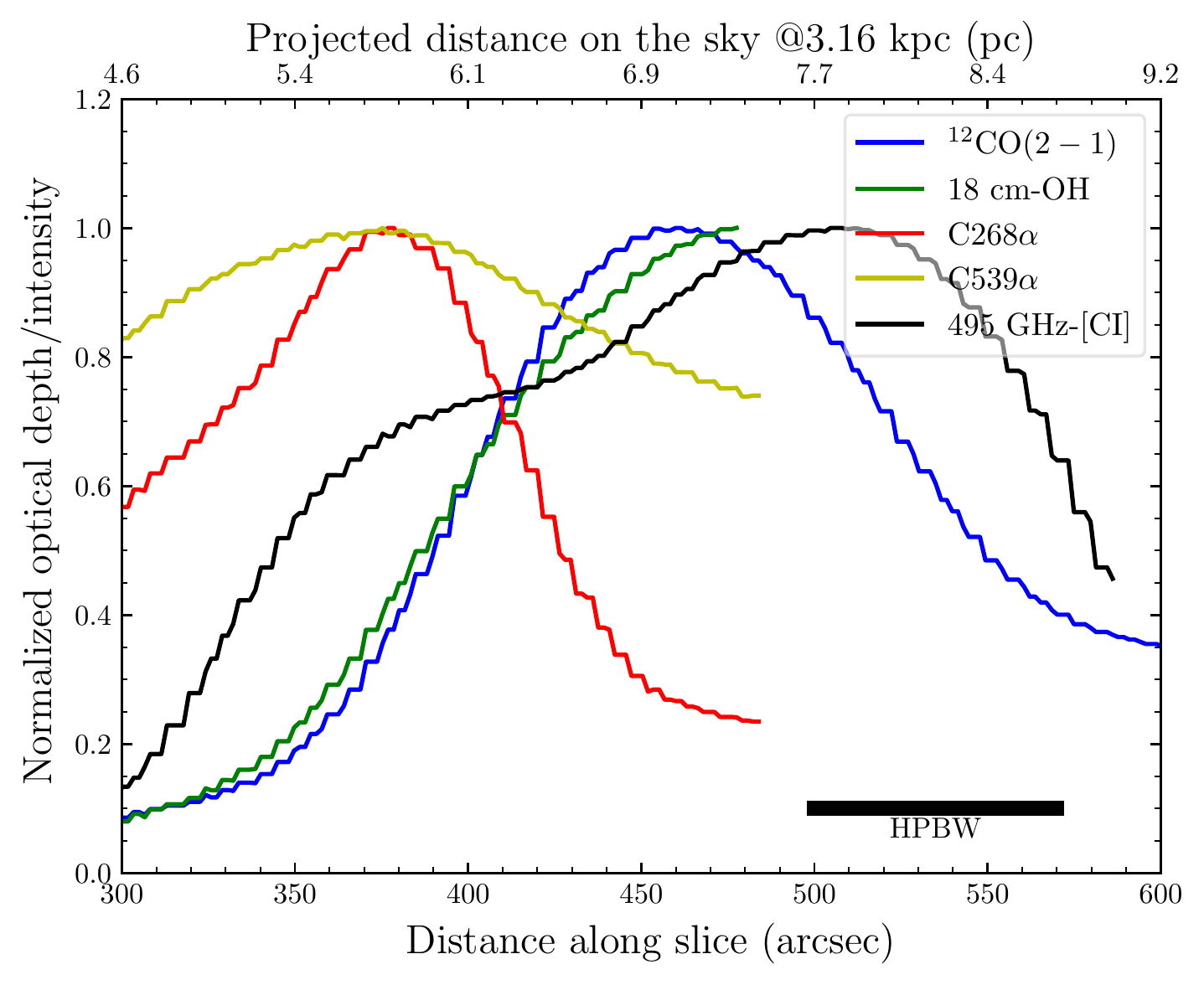}
 \end{center}
 \caption{Comparison between different ISM tracers along a slice joining the C$268\alpha$ optical depth peak and $^{12}$CO$(2$--$1)$ line peak at a velocity of $-47.8$~km~s$^{-1}$.
 The {\it black bar} at the bottom shows the half power beam width ($70\arcsec$) of the images.
 The {\it top axis} shows the plane of the sky distance along the slice assuming that the gas is at a distance of $3.16$~kpc from the observer \citep{Salas2017}.
 Given that the gas could be closer to the observer this distance is an upper limit.}
 \label{fig:slice1}
\end{figure}

To explore the relation between CO emission and CRRL emission we draw a slice joining the peaks of $^{12}$CO$(2$--$1)$ and C$268\alpha$ emission at a velocity of $-47.8$~km~s$^{-1}$.
The slice is shown as a green line in Figure~\ref{fig:hicrrlco} in the panels with a velocity of $-47.8$~km~s$^{-1}$.
The normalised intensity or optical depth of different tracers along this slice is shown in Figure~\ref{fig:slice1}.
Here we notice how the optical depth of C$268\alpha$ and C$\alpha(537)$ peaks at the same location, and the molecular lines peak towards the left of the CRRLs, which corresponds to the South-East direction in the sky.
The difference between the peaks of the CRRLs and the molecular lines is similar to that expected in a photo-dissociation region \citep[PDR, e.g.,][]{Hollenbach1999}.
As we move towards the South-East of Cas~A the gas shows a CII/CI/CO layered structure, which suggests that we are observing the photodissociation region associated with the edge of a molecular cloud.

The distance at which the gas becomes CO bright will depend on the average PDR density.
The projected distance on the plane of the sky between the peak of the C$268\alpha$ optical depth and the peak of the $^{12}$CO$(2$--$1)$ emission is $1.3\pm0.6\arcmin$.
If we assume that the Perseus arm gas is at a distance of $3.16\pm0.02$~kpc from Earth in the direction of Cas~A \citep{Choi2014,Salas2017}, then this corresponds to $\sim1.2\pm0.5$~pc in the plane of the sky.
CO will be sufficiently shielded from photo-dissociating photons when $A_{\rm{FUV}}\sim1$.
We adopt a conversion factor between extinction in the V band and hydrogen column density of $N_{\rm{H}}=(2.08\pm0.02)\times10^{21}A_{\rm{V}}$~cm$^{-2}$ \citep{Zhu2017}.
Then, to convert between optical opacity and far-ultra violet (FUV) opacity we adopt $\kappa_{\rm{d}}(\rm{FUV})\approx1.8\kappa_{\rm{d}}(\rm{V})$.
With this, for an $A_{\rm{FUV}}$ of one magnitude we have $N_{\rm{H}}=(1.15\pm0.01)\times10^{21}$~cm$^{-2}$.
This implies that the mean density in this PDR is $n_{\rm{H}}=310\pm28$~cm$^{-3}$. 
This density is consistent with the hydrogen density derived from the CRRL analysis (Table~\ref{tab:gasprops}).

Motivated by the observed layered structure we compare the CO emission to that of an edge-on PDR model.
This model is an extension of the \citet{Tielens1985a} PDR model which includes the updates of \citet{Wolfire2010} and \citet{Hollenbach2012}.
The calculation of line intensities and source parameters for edge-on models are discussed in \citet{Pabst2017}.
We use a total hydrogen density of $300$~cm$^{-3}$, an $A_{\rm{V}}$ of eight along the line of sight and of eight in the transverse direction.
The gas in the PDR is illuminated on one side by an interstellar radiation field with $G_{0}=4.2$, measured in \citet{Habing1968} units, and primary cosmic ray ionisation rate per hydrogen of $7\times10^{-17}$~s$^{-1}$.
The carbon and hydrogen RRLs observed towards Cas~A \citep{Oonk2017} have been reanalysed by \citet{Neufeld2017b} taking into account the relevant chemical recombination routes and we have adopted values of the radiation field and cosmic-ray ionisation rate consistent with their results.
We adopt an abundance $^{12}$CO/$^{13}$CO of $60$, appropriate for gas in the Perseus arm in this direction \citep{Langer1990,Milam2005}.

\begin{figure}
 \begin{center}
  \includegraphics[width=0.49\textwidth]{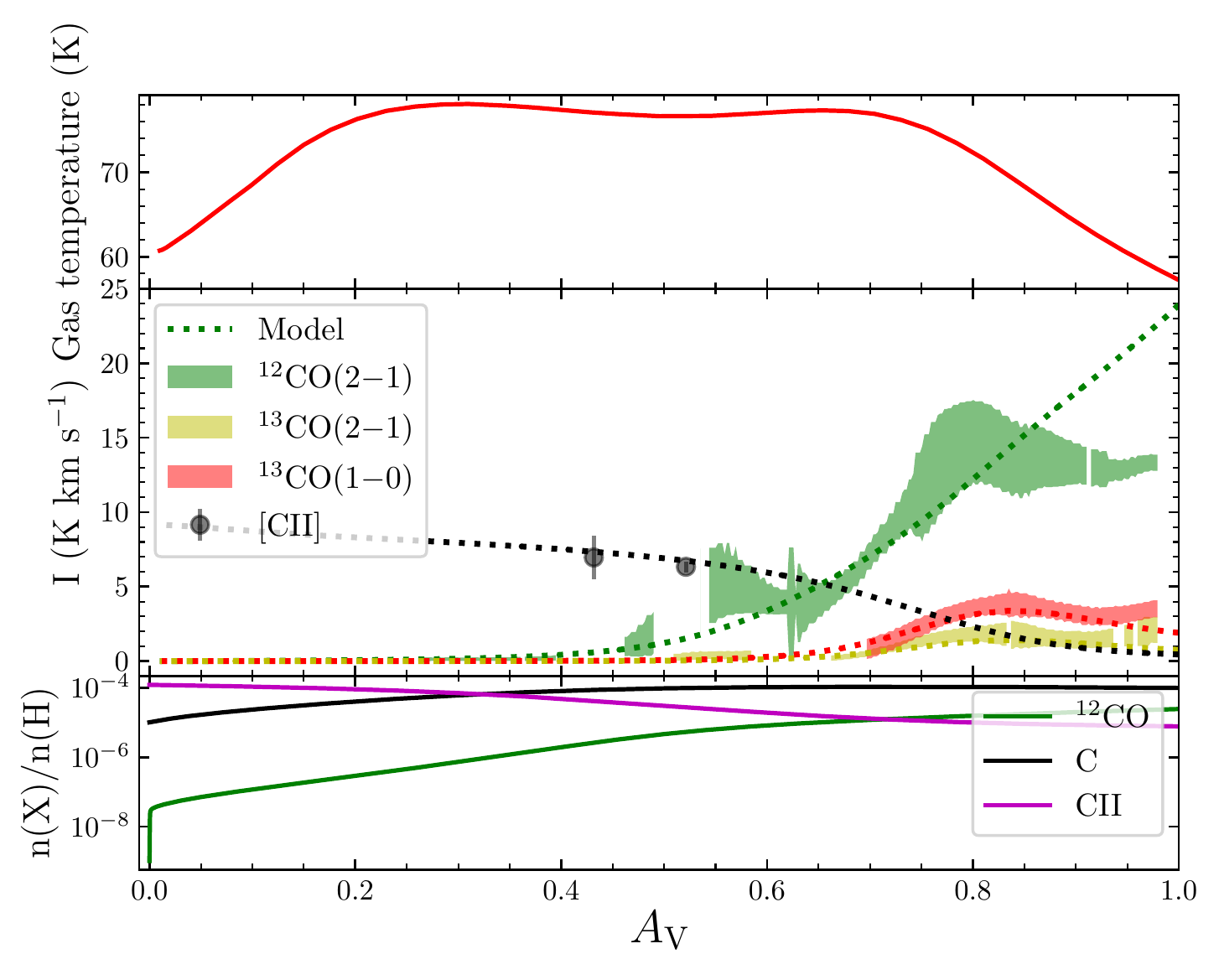}
 \end{center}
 \caption{Comparison between the $^{12}$CO$(2$--$1)$, $^{13}$CO$(1$--$0)$, $^{13}$CO$(2$--$1)$ and $158$~$\mu$m [CII] velocity integrated line intensity along the slice shown in Figure~\ref{fig:slice1}.
 The observed quantities ({\it shaded regions}) are compared with a PDR model ({\it dashed lines}).
 The {\it shaded regions} show the observed values and their $3\sigma$ error range.
 The $158$~$\mu$m [CII] points have been rescaled to be visible in this Figure.
 The PDR model is that of a region with a constant hydrogen density of $300$~cm$^{-3}$, an incident radiation field of $G_{0}=4.2$ on one side of the cloud, and a line of sight optical interstellar extinction of eight}.
 For more details on the PDR model see the text and \citet{Pabst2017}.
 \label{fig:slicecomp}
\end{figure}

The comparison between observations and the output from the PDR model is presented in Figure~\ref{fig:slicecomp}.
The adopted density ensures that the calculated distance on the sky between the CO peak and the surface of the PDR, as defined by the CRRL peak, agrees with the observations.
For $A_{\rm{V}}\geq0.5$ the increase in the line intensity is well described by the model. 
For $A_{\rm{V}}\sim0.9$ the proximity to the edge of the mapped region causes the velocity integrated line intensity to decrease.
This decrease close to the map edge is caused by the convolution with a $70\arcsec$ beam.

Additionally, we use the same model to predict the velocity integrated line intensity of the $492$~GHz-[CI] and $158$~$\mu$m-[CII] lines, and the optical depth of the $18$~cm-OH line.
The model does a good job in reproducing the observed optical depth of the OH line.
In the region where the slice intersects footprint $1$ of the PACS $158$~$\mu$m-[CII] cube (Figure~\ref{fig:ciimap}), the model predicts a value of $4.3\times10^{-5}$~erg~cm$^{-2}$~s$^{-1}$~sr$^{-1}$.
The observed value is $56\%$ larger, which can be accounted for by the prescence of gas at higher velocities not present in the model (e.g., the $-38$ and $0$~km~s$^{-1}$ velocity components present in the velocity unresolved PACS observations).
However, in the case of atomic carbon, the model overestimates the observed values by a factor of five.
This is similar to that found in other lines of sight, where the predicted atomic carbon column density is larger than the observed one \citep[e.g.,][]{Gong2017}.

\citet{deLooze2017} find an interstellar radiation field (ISRF) with a strength $G_{0}\sim0.6$.
This value is lower than the one adopted here, but we do note that against Cas~A it is not possible to use the dust spectral energy distribution to estimate the strength of the ISRF.
Outside the area covered by Cas~A \citet{deLooze2017} find strengths for the ISRF of the order of unity.
However, the derived strength of the ISRF will depend on the adopted model with a variation of up to $1.6$ depending on the model details \citep[e.g.,][]{Fanciullo2015,PlanckXXIX2016}.
In their work \citet{deLooze2017} also use line ratios and the PDR toolbox models \citep{Pound2008} to estimate the strength of the ISRF in the ISM between Cas~A and Earth.
They find that the line ratios are consistent with their dust derived value of $G_{0}\sim0.6$, but the line ratio is also consistent with a lower density and stronger ISRF \citep[see Figure~C1 in][]{deLooze2017}.
Based on the current data we infer that the adopted value is in reasonable agreement with all observations.

\subsubsection{Gas column density}

For the $-47$~km~s$^{-1}$ feature most of the $21$~cm-HI optical depth maps show that the line is saturated with values of $\tau\gtrsim5$ \citep{Bieging1991}.
Nonetheless, this lower limit on the optical depth can be used to place a lower limit on the atomic hydrogen column density.
If we assume that the width of the $21$~cm-HI line profile at $-47$~km~s$^{-1}$ is the same as that of the CRRL at the same velocity, and that the spin temperature is greater than $50$~K, then we have that $N(\rm{HI})>1.5\times10^{21}$~cm$^{-2}$.
This limit is consistent with the column density derived from the CRRL and edge-on PDR analysis and it implies a fraction $N(\rm{HI})/N(\rm{H_{2}})\gtrsim0.1$.

Additional estimates of the gas column density can be obtained from measurements of X-ray absorption and from the dust optical depth.
In the case of X-ray absorption \citet{Hwang2012} determined values of $2\times10^{22}$~cm$^{-2}$ over the South portion of Cas~A, with higher values ($\sim3\times10^{22}$~cm$^{-2}$) towards its western hotspot.
These values are slightly smaller than the ones found using the CRRLs lines.
We consider that this difference is not significant given the uncertainties associated with X-ray column density measurements \mbox{\citep[e.g.,][]{Predehl1995,Zhu2017}}.
Recently, \citet{deLooze2017} modelled the dust emission towards Cas~A and used it to determine the mass of dust in the ISM along the line of sight.
They adopted a dust-to-gas ratio of $0.0056$ and found column densities of $1.5\times10^{22}$~cm$^{-2}$ towards the south of Cas~A, and $2.2\times10^{22}$~cm$^{-2}$ towards the western hotspot.
A map showing the spatial distribution of column density derived from the dust analysis is shown in the rightmost panel of Figure~\ref{fig:crrlemelnc}.
A comparison between the column densities derived from the dust and CRRL analysis (right panels in Figure~\ref{fig:crrlemelnc}) shows good agreement, with larger values towards the South of Cas~A and a peak against its western hotspot.
To compare their magnitudes we focus on regions towards the South of the centre of Cas~A, where we see less emission from gas at $-38$~km~s$^{-1}$ which would create confusion.
Here the magnitude of the CRRL derived gas column density, $(2.4\mbox{--}11.3\times10^{22}$~cm$^{-2}$, is comparable to that derived from the dust analysis.
The major uncertainty in the determination of the ISM dust content along this line of sight comes from the separation between the foreground ISM dust component and the contribution from dust associated with the supernova remnant.
This introduces a factor of a few uncertainty in the derived dust mass and column density.

The CRRL derived gas column density averaged over the face of Cas~A, $(3.4$--$25.9)\times10^{18}$~cm$^{-2}$, on its own implies an hydrogen column density of $(2.3\mbox{--}17.3)\times10^{22}$~cm$^{-2}$.
Given that not all of the carbon is ionized, we also need to account for carbon in atomic and molecular forms.
We focus on a region to the South of the center of Cas~A, following the slice in Figure~\ref{fig:hicrrlco}, where there is CO emission.
Here, with an adopted $A_{\rm{V}}$ of eight, our PDR model is able to reproduce the CO observations.
However, for an $A_{\rm{V}}$ of eight and an excitation temperature $\sim20$~K, the CO lines used here are optically thick.
This points towards the presence of denser CO-rich clumps embedded in a lower density CO-dark halo.
This situation is similar to that observed towards the W43 star forming region, where large column densities are derived from atomic hydrogen observations at $21$~cm \citep[$N_{\rm{H}}\approx2\times10^{22}$~cm$^{-2}$,][]{Bihr2015,Bialy2017}.
Gamma-ray and dust observations in our Galaxy have revealed the prescence of large reservoirs of CO-dark molecular gas \citep[with a mass fraction comparable to that of the CO molecular gas][]{Grenier2005,PlanckXIX2011}.
Likely, much of this gas is in extended atomic hydrogen halos around giant molecular cloud complexes in spiral arms such as the ones probed in this study in the Perseus arm.

\subsection{Envelopes of molecular clouds}

The gas properties derived from the CRRL analysis seem to bridge the gap between the atomic gas traced by the $21$~cm line of HI and the molecular gas traced by the CO lines in the millimimeter \citep{Oonk2017,Salas2017}.
From spatially unresolved observations of the $21$~cm-HI line in this direction \citet{Davies1972} derived a temperature of $140\pm40$~K for the two most prominent Perseus arm absorption features at $-47$ and $-38$~km~s$^{-1}$.
It has not been possible to estimate this value on smaller scales due to the saturation of the $21$~cm-HI line at these velocities \citep[e.g.,][]{Bieging1991}, but it is likely to be slightly colder.
On the molecular side of things we have temperatures of $\sim20$~K \citep{Batrla1984}.
This would put the gas traced by low frequency CRRLs, with an electron temperature of $\sim80$~K, in between atomic and molecular gas.
A place where this transition takes place is the envelope of molecular clouds \citep[e.g.,][]{Moriarty-Schieven1997,Krumholz2009,Sternberg2014}.

Studies of molecular clouds in the solar vicinity show that their atomic envelopes have temperatures of the order $\sim100$~K \citep{Andersson1991}.
A comparisson between gas traced by the $21$~cm HI and the $^{13}$CO lines shows that the atomic component is more extended than the molecular one and their velocity fields are not necessarily aligned \citep{Imara2011}.
The properties of the envelope will depend on the environment.
Here we compare against two giant molecular clouds, both show large fractions of atomic gas, but one shows little star formation, with an infrared luminosity of $\approx5000$~L$_{
\odot}$; G216--2.5 \citep[the Maddalena-Thaddeus cloud,][]{Maddalena1985,Williams1996,Megeath2009,Imara2015} and the other a mini-starburst with an infrared luminosity of $3.5\times10^{6}$~L$_{\odot}$ \citep{Motte2003,Nguyen-Luong2011b,Bihr2015,Bialy2017}.
The atomic envelope around G216--2.5 has a thickness of $\sim50$~pc and the atomic gas column density inferred from observations of the $21$~cm line of HI is $\approx2.21\times10^{21}$~cm$^{-2}$ \citep{Williams1996}.
Around W43, the atomic envelope has a thickness of $140$~pc \citep{Motte2014}, and the atomic gas column density is $\approx2\times10^{22}$~cm$^{-2}$ \citep{Bihr2015,Bialy2017}.
The later column density is consistent with the lower end of the ranges found here (Table~\ref{tab:gasprops}).

Another way in which we can study the envelopes of giant molecular clouds is with observations of the $158$~$\mu$m-[CII] line.
In cases where it is possible to isolate an ionized carbon layer around a molecular cloud it is found that the gas temperature and density are close to those found here.
In the Magellanic clouds \citet{Pineda2017} find densities of $700$--$2000$~cm$^{-3}$ and temperatures of $\sim50$~K.

\subsection{Uncertainties in the CRRL modelling}

The change in line properties as a function of physical properties is quite sensitive to the gas physical conditions.
During the modelling of the CRRL line properties as a function of principal quantum number a series of assumptions are made which have an effect on the derived gas properties.
Some of these have been explored previously, like the use of different angular momentum changing collisional rates \citep{Salgado2017a}, or including collisions with hydrogen when solving the level population problem \citep{Oonk2017}.
Additionally, two assumptions have not been explored before, those are; the use of different collisional rates for the excitation of ionized Carbon and changing the carbon and electron abundances relative to hydrogen.
Having different collisional rates and abundances will change the dielectronic capture rate.
This will be reflected as a change in the departure coefficient $b_{n}$, which determines how the integrated optical depth will behave as a function of $n$.

\begin{figure}
 \begin{center}
  \includegraphics[width=0.49\textwidth]{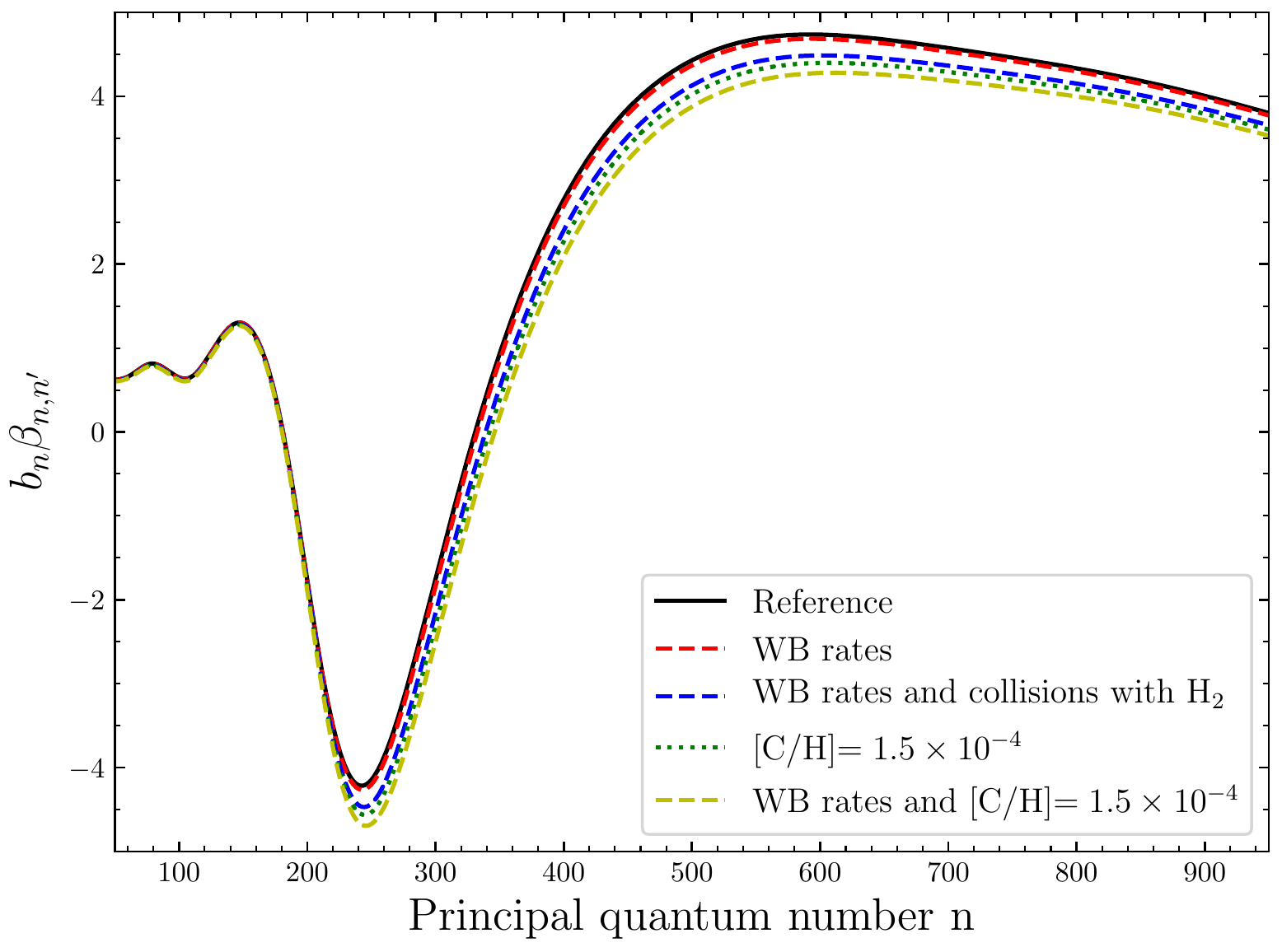}
 \end{center}
 \caption{Comparison between CRRL models under different assumptions regarding the carbon ion collisional excitation rates, carbon and electron abundances.
 All models were computed for an electron density of $0.05$~cm$^{-3}$ and temperature $80$~K.
 The reference model is one computed assuming $[\rm{C}/\rm{H}]=[\rm{e}/\rm{H}]=3\times10^{-4}$ and using the \citet{Tielens1985a} and \citet{Hayes1984} collisional rates.
 The models labelled with WB rates use the \citet{Wilson2002} rates for collisions with electrons and those of \citet{Barinovs2005} for collisions with atomic hydrogen.
 The model which includes collisions with molecular hydrogen, H$_{2}$, uses the rate for collisions with molecular hydrogen presented in \citet{Goldsmith2012}.
 The models labelled with $[\rm{C}/\rm{H}]=1.5\times10^{-4}$ use this abundance for carbon and electrons.
 The largest difference for and individual change is produced when the electron and carbon abundances are modified.
 }
 \label{fig:crrlmods}
\end{figure}

Given that it is computationally expensive to recompute the grid of models for each set of assumptions, we focus on a particular point in the $n_{\rm{e}}-T_{\rm{e}}$ plane.
The change in the $b_{n}\beta_{n,n^{\prime}}$ coefficients as a function of $n$ is shown in Figure~\ref{fig:crrlmods} for different assumptions.
These show that the change in the integrated optical depth will be $\sim10\%$ for the models computed using the different collisional rates and lower carbon and electron abundances.
For other assumptions the difference will be lower.

Recently, \citet{Guzman2017} and \citet{Vrinceanu2017} have investigated the effect of using different formulations (semi-classical versus quantum mechanical) when computing the $\ell$-changing collisional rates.
As shown by \citet[][See their Figure~14]{Salgado2017a} this will affect the departure coefficients, mainly their absolute values.
However, as \citet{Guzman2017} and \citet{Vrinceanu2017} point out, there is no physical reason to prefer one formulation over the other.
A more detailed comparison of the effect different formulations will have on the predicted CRRL properties will be investigated in the future.

\section{Summary}
\label{sec:summary}

We have presented $70\arcsec$ resolution CRRL maps at $340$, $148$, $54$ and $43$~MHz.
The distribution of the C$268\alpha$ line in emission reveals a good correlation with regions where the $21$~cm-HI line is saturated and regions of faint $^{12}$CO$(2\mbox{--}1)$ emission.
We interpret this as as a diffuse PDR, in which low frequency CRRLs trace the less dense, warmer envelope of molecular gas.

Using the ratios between CRRLs we have constrained the gas electron temperature and density along the line of sight on scales of $\lesssim1.2$~pc.
With the line ratios used here, the constraints result in a range of allowed electron temperatures and densities.
Averaged over the face of Cas~A the constraints on the electron density are $n_{\rm{e}}=(0.03\mbox{--}0.055)$~cm$^{-3}$ and $T_{\rm{e}}=(70\mbox{--}140$~K for gas in the Perseus arm of the Galaxy at $-47$~km~s$^{-1}$.
The pressure shows variations of less than a factor of two on $\lesssim1.2$~pc scales.

From the constraints on the electron temperature and density we derived lower limits for the ionized carbon emission measure, column density and its line of sight path length.
The lower limit on the column density is $3.4\times10^{18}$~cm$^{-2}$, which corresponds to an hydrogen column density of $2.2\times10^{22}$~cm$^{-2}$ if all carbon is ionized and [C/H]$=1.5\times10^{-4}$.
The hydrogen density derived from analysis of the CRRLs integrated optical depths is $210$--$360$~cm$^{-3}$.

A PDR model with an average hydrogen density of $300$~cm$^{-3}$, and conditions similar to those inferred for the clouds in this region, is able to reproduce the observed distribution of $^{12}$CO$(2\mbox{--}1)$, $^{13}$CO$(2\mbox{--}1)$, $^{13}$CO$(1\mbox{--}0)$ and $1667$~MHz OH.

The relatively high spatial resolution of the present observations enables us to study the relation between CRRLs and other tracers of the ISM on scales where it is possible to observe the PDR like structure in the surface of a molecular cloud.
This also highlights the importance of CRRLs as tracers of the diffuse ISM, as they allow us to determine the gas physical conditions in regions which are not readily traced by $21$~cm-HI and/or CO.
These observations highlight the utility of CRRLs as tracers of low density extended HI and CO-dark gas halo's around molecular clouds.
Future surveys of CRRLs with the low frequency array (LOFAR) are promising as they could reveal important new clues about the physics of the ISM, particularly about the transition from atomic-to-molecular gas and the properties of CO-dark gas.

\section*{Acknowledgements}
{We would like to thank the anonymous referee for useful comments.
P.~S., J.~B.~R.~O., A.~G.~G.~M.~T, H.~J.~A.~R. and K.~L.~E. acknowledge financial support from the Dutch Science Organisation (NWO) through TOP grant 614.001.351.
LOFAR, designed and constructed by ASTRON, has facilities in several countries, that are owned by various parties (each with their own funding sources), and that are collectively operated by the International LOFAR Telescope (ILT) foundation under a joint scientific policy.
We gratefully acknowledge that LCASS is carried out using Directors discretionary time under project DDT001.
M.~C.~T. acknowledges financial support from the NWO through funding of Allegro.
M.~G.~W. was supported in part by NSF grant AST1411827.
A.~G.~G.~M.~T acknowledges support through the Spinoza premie of the NWO.
This research made use of Astropy, a community-developed core Python package for Astronomy \citep{Astropy2013}, and of NASA's Astrophysics Data System.
The LOFAR software and dedicated reduction packages on https://github.com/apmechev/GRID\_LRT were deployed on the e-infrastructure by the LOFAR e-infragroup, consisting of J.~B.~R.~Oonk (ASTRON \& Leiden Observatory), A.~P.~Mechev (Leiden  Observatory) and T.~Shimwell (Leiden Observatory) with support from N.~Danezi (SURFsara) and C.~Schrijvers (SURFsara). This work has made use of the Dutch national e-infrastructure with the support of SURF Cooperative through grant e-infra160022.
}

{\it Facilities:} WSRT, LOFAR, Herschel, VLA.

\newpage
\bibliographystyle{mnras}
\bibliography{ref_rrl.bib}

\bsp    
\label{lastpage}
\end{document}